\documentclass[a4paper,11pt]{article}
\usepackage[english]{babel}
\usepackage[latin1]{inputenc}
\usepackage[T1]{fontenc} 
\usepackage{ae,aecompl} 
\usepackage[strict]{changepage}
\usepackage{amssymb}
\usepackage{amsmath}
\usepackage{graphicx}
\usepackage{subfig}
\usepackage{fancyhdr}
\IfFileExists{.pdfmode}{%
  
  \usepackage{ae,aecompl} 
  }{
  
  }

\begin{document}

\thispagestyle{plain}
\begin{center}
\Huge{Trojan resonant dynamics,}\\
\Huge{ stability, and chaotic diffusion}\\
\huge{for parameters relevant to exoplanetary systems}
\end{center}

\begin{center}
\Large{Rocio Isabel P\'aez}\\
\small{Dip. di Matematica, Universit\`a di Roma ``Tor Vergata'', Italy}\\
\small{\textsc{paez@mat.uniroma2.it}}\\
\vspace{0.3cm}
\Large{Christos Efthymiopoulos}\\
\small{Research Center for Astronomy and Applied Mathematics, Academy of Athens, Greece}\\
\small{\textsc{cefthim@academyofathen.gr}}\\
\end{center}

\noindent
{\bf Abstract:} The possibility that giant extrasolar planets 
could have small trojan co-orbital companions has been examined in the
literature from both viewpoints of the origin and dynamical stability
of such a configuration. Here we aim to investigate the dynamics 
of hypothetical small trojan exoplanets in domains of secondary
resonances embedded within the tadpole domain of motion. To this 
end, we consider the limit of a massless trojan companion of a giant 
planet. Without other planets, this is a case of the elliptic restricted 
three body problem (ERTBP). The presence of additional planets (hereafter 
referred to as the restricted multi-planet problem, RMPP) induces new 
direct and indirect secular effects on the dynamics of the trojan body. 
The paper contains a theoretical and a numerical part. In the theoretical 
part, we develop a Hamiltonian formalism in action-angle variables, 
which allows to treat in a unified way resonant dynamics and 
secular effects on the trojan body in both the ERTBP or the RMPP. 
In both cases, our formalism leads to a decomposition of the Hamiltonian 
in two parts, $H=H_b+H_{sec}$. $H_b$, called the basic model, describes 
resonant dynamics in the short-period (epicyclic) and synodic (libration) 
degrees of freedom, while $H_{sec}$ contains only terms depending 
trigonometrically on slow (secular) angles. $H_b$ is formally identical 
in the ERTBP and the RMPP, apart from a re-definition of some angular 
variables. An important physical consequence of this analysis is that 
the slow chaotic diffusion along resonances proceeds in both the ERTBP 
and the RMPP by a qualitatively similar dynamical mechanism. We found 
that this is best approximated by the paradigm of `modulational diffusion'. 
In the paper's numerical part, we then focus on the ERTBP in order to 
make a detailed numerical demonstration of the chaotic diffusion 
process along resonances. Using color stability maps, we 
first provide a survey of the resonant web for characteristic 
mass parameter values of the primary, in which the secondary resonances 
from 1:5 to 1:12 (ratio of the short over the synodic period), as 
well as their transverse resonant multiplets, appear. We give numerical 
examples of diffusion of weakly chaotic orbits in the resonant web. 
We finally make a statistics of the escaping times in the resonant 
domain, and find power-law tails of the distribution of the escaping 
times for the slowly diffusing chaotic orbits. Implications of resonant 
dynamics in the search for trojan exoplanets are discussed.

\begin{center}
\line(1,0){250}
\end{center}

\section{Introduction}

A number of studies (Menou and Tabachnik 2003, Dvorak et al. 2004,
\'{E}rdi and S\'{a}ndor 2005, Beaug\'{e} et al. 2007, Schwarz et
al. 2007, Lyra et al. 2009, Cresswell and Nelson 2009, Funk et
al. 2012) have pointed toward the hypothesis that small (at most
Earth-sized) planets could exist as Trojan companions of giant planets
in extrasolar planetary systems. The possibility that pairs of planets
might be found in co-orbital resonance in general was first
investigated by Laughlin and Chambers (2002). For current and future
possibilities to observationally detect extrasolar trojan planets see
Giuppone et al. (2012), Haghighipour (2013), Dobrovolskis (2013), and
references therein.

In order to ensure that a hypothetical small trojan planet could 
exist close to a giant planet up to the lifetime of its hosting system, 
of major interest becomes the question regarding the {\it long-term stability} 
of trojan motions in various models of extrasolar planetary systems. 
Computing the size and form of the domain of stability in the framework 
of the circular or elliptic restricted three-body problem is a classical 
problem that has attracted study since decades (see e.g. Rabe 1967, 
Deprit 1966, Garfinkel 1977, \'{Erdi} 1988, Celletti and Giorgilli 1991,
Lohinger and Dvorak 1993, \'{E}rdi 1997, Giorgilli and Skokos 1997,
Gabern et al. 2005, Efthymiopoulos and S\'{a}ndor 2005, Lhotka et al.
2008, Efthymiopoulos 2013). As exemplified in dynamical studies of
Jupiter's trojan asteroids in our own solar system (Levison et
al. 1997, Marzari et al. 2003, Tsiganis et al. 2005, Robutel et
al. 2005, Robutel and Gabern 2006, Robutel and Bodossian 2009), a key
role in the problem of long term stability is played by the
co-existence of various types of {\it resonances} embedded in the
co-orbital domain . Besides the so-called `secondary resonances'
(\'{E}rdi et al 2007, 2009), i.e. resonances involving the short and
long periods of the trojan body (see subsection 2.4 for precise 
definitions), there are also resonances involving
the secular periods of either the trojan body or of its giant
companion and/or other giant planets in the same system (see Robutel
2005, Robutel and Gabern 2006). Finally, resonances involving some
`great inequality'-type motion between the giant planets may
occasionally penetrate a trojan swarm, in particular during the early
stages of planetary migration (Robutel and Bodossian 2009). The main
effect of all these resonant interactions is to produce domains of
chaotic diffusion in phase space. Main questions are: i) the geometry
and the extent of resonant domains, ii) the speed of chaotic
diffusion, and iii) the statistics of the thereby induced times of
escape.

Since in observed extrasolar planetary systems gaseous exoplanets
appear in an extended range of mass parameters and eccentricities, a
parametric study of the effects of resonances on co-orbital dynamics
appears necessary.  Such a study, aiming to determine the size of the
stability domain as a function of the gaseous planet's mass parameter
$\mu$, was performed by \'{E}rdi et al. (2007, 2009) in the framework 
of the elliptic restricted three body problem (ERTBP). These studies 
practically measure how the size of the main stable domain of the phase 
space around $L_4$ and $L_5$ is affected by the presence of secondary 
resonances. However, for $\mu$ in the range $0.001\leq \mu\leq 0.01$,
we encounter quite {\it low} order resonances of this type,
i.e. resonances of the type 1:n between the short and long period
with a relatively small integer $n$, i.e., $12\geq n\geq 4$. Hence the
phase space volume occupied by the resonances themselves becomes important. 
Furthermore, as shown in section 3 below, the resonant multiplets formed 
around a low-order secondary resonance, due to the effects of secular
frequencies, consist of groups of resonances whose separation in action
space is neither very small (as e.g. for the trojan asteroids in
our solar system), nor very big (so as to completely avoid resonance
overlap).  This `intermediate' regime regarding the separation of
resonances in a multiplet is due to the fact that, as $\mu$ increases,
the timescales associated with the three main types of motion, i.e.,
short-period (of frequency ${\cal O}(1)$), long-period (of frequency
${\cal O}(\sqrt{\mu})$) and secular (of frequency ${\cal O}(\mu_j)$, 
where $\mu_j$ is the mass parameter of the j-th planet in the extrasolar 
system), no longer separate very well one from the other.  This, in turn, 
complicates the question of characterizing the diffusion within the 
resonant domain according to well known paradigms of dynamics 
(e.g.  `Arnold', modulational, or resonance overlap (Chirikov) 
diffusion).

In the present paper, we initiate a parametric study of the resonant
structure in the trojan co-orbital domain, as well as of the main
types and associated timescales for chaotic diffusion of hypothetical
small (considered massless) trojan exoplanets. The paper 
contains a theoretical and a numerical part. Our numerical study is 
in the framework of the planar ERTBP. However, in section 2 we introduce 
a formalism, based on action-angle variables, which renders straightforward 
generalizations including secular effects of more than one exoplanets. 
In our study we focus on mass parameters in the range $0.001\leq \mu\leq 
0.006$ for the giant primary. This includes the secondary resonances 
around the ratios (of the short to long period) from 1:12 to 1:5. 
For still higher $\mu$ (around 0.01) there appear the resonances 
1:4 and 1:3, which however, present particularities rendering necessary 
a separate study (see Deprit 1966, Deprit and Price 1969, Deprit and 
Henrard 1969, Dullin and Worthington 2013).

The structure of the paper is as follows: section 2 contains theoretical 
considerations and introduces convenient action-angle variables for our 
study. We demonstrate that, assuming a multi-planet system far 
from mean motion resonances, the Hamiltonian can be given the form 
$H=H_b+H_{sec}$, where: i) $H_b$, called the `basic model', is a model 
of two degrees of freedom involving two action-angle pairs that characterize 
the short period and the synodic motions of the trojan body respectively, 
ii) $H_{sec}$ contains only `slow' terms, i.e., trigonometric terms depending 
on secular frequencies. Hence, the dynamics at resonances is approximated 
as the one due to $H_b$ slowly modulated by the additional influence of 
$H_{sec}$. Furthermore, with a re-interpretation of variables, the form 
of $H_b$ is identical in the RMPP and the ERTBP. Finally, we show how 
to use these variables in order to compute proper elements for resonant 
trojan motions. Section 3 is devoted to a parametric study of the stability 
maps and phase portraits for various values of $\mu$ and $e'$ 
(the eccentricity of the giant primary), in the framework of the 
ERTBP, in which, our characterization of resonances stems from the 
general decomposition of the Hamiltonian as in section 2. Using a 
plane corresponding to our definition of proper elements, we compute 
and depict the form of the corresponding web or resonances in the 
plane of proper elements, whereby the volume and relative importance 
of each resonance can be estimated. Section 4 presents some main types 
of chaotic diffusion observed by numerically computing orbits with 
initial conditions in properly choosen chaotic subsets of the resonance 
web. We provide some criteria to characterize the chaotic diffusion. 
Finally, we integrate ensembles of orbits and make a statistical study 
of the escape probabilities as well as the distribution of escape times. 
Section 5 summarizes the main conclusions of the present study.

\section{Theory}

\subsection{Hamiltonian formalism}
The Hamiltonian of the planar ERTBP can be written in heliocentric
canonical variables $(\mathbf{r},\mathbf{p})$ as:
\begin{equation}\label{ham1}
H={p^2\over 2} -{G M\over r} - G m' \left({1\over\Delta}
-{\mathbf{r}\cdot\mathbf{r}'\over r'^3}\right)
\end{equation}
where $m'$ is the mass of the primary planet, $M$ the mass of the star,
$\mathbf{r}$, $\mathbf{r'}$ the heliocentric position vectors of the
test particle and of the primary respectively, $r=|\mathbf{r}|$,
$r'=|\mathbf{r'}|$ and $\Delta = \vert \mathbf{r} - \mathbf{r}' \vert$.

The primary's major semi-axis, eccentricity, and argument of the
perihelion are denoted by $a',e',\omega'$ respectively. We set the
unit of length as $a'=1$, and the unit of time such that the mean
motion of the primary is equal to $n=1$. This implies $G(M+m')=1$. 
Defining the mass parameter as $\mu=G m'$, the Hamiltonian
in the above units takes the form:
\begin{equation}\label{ham2}
H={p^2\over 2} -{1\over r} - \mu \left({1\over\Delta}
-{\mathbf{r}\cdot\mathbf{r}'\over r'^3}-{1\over r}\right)~~.
\end{equation}

We now introduce modified Delaunay action - angle variables in the
above model. As in Brown and Shook (1962; see also Morais 2001), 
including the small Keplerian correction $\mu/r$ in the disturbing 
function allows to define Delaunay variables with values independent
of $\mu$:
\begin{equation}\label{del}
x=\sqrt{a}-1,~~~y=\sqrt{a}\left(\sqrt{1-e^2}-1\right)
\end{equation}
where $(a,e)$ are the major semi-axis and the eccentricity. 
The action variables $(x,y)$ have the conjugate angles $(\lambda,\omega)$, 
i.e. the mean longitude and argument of the perihelion of the trojan 
(massless) body respectively.  The Hamiltonian now reads
\begin{equation}\label{ham3}
H=-{1\over 2(1+x)^2} - \mu R(\lambda,\omega,x,y,
\lambda'=n t;\omega',e',a'=1)~~.
\end{equation}
As shown below, in computing proper elements it turns to be 
crucial to formally remove the explicit dependence of the Hamiltonian 
on time by introducing a `dummy' action variable $I_3$ conjugate to
$\lambda'$, namely
\begin{equation}\label{ham4}
H=-{1\over 2(1+x)^2} + I_3
- \mu R(\lambda,\omega,x,y,\lambda';\omega',e')~~.
\end{equation}

The Hamiltonian (\ref{ham4}) represents a system of three degrees of
freedom. We now augment the number of degrees of freedom by 
including the secular effects of additional planets. 
To this end, assuming that the planets are far from mean motion 
resonances, it is convenient to approximate the secular dynamics 
as quasi-periodic, thus introducing a set of secular frequencies,  
$g'$ for the primary, and $g_1,g_2,...g_S$ for $S$ additional planets. 
The frequencies $g',g_j,j=1,\ldots,S$ can be computed either by a linear 
(Laplace) theory, or by a non-linear analytical extension (see, for example, 
Libert and Sansottera 2013, and references therein), or by purely numerical 
methods (e.g. frequency analysis, see Laskar 2004). In either case, we adopt 
the convention that the frequencies $g',g_j$ are the frequencies of the 
leading terms in the quasi-periodic representation of the oscillations of 
the planets' eccentricity vectors. Thus, the time evolution of the latter
can be approximated by the equations (see, for example, Morbidelli, 
2002, chapter 7)
\begin{equation}\label{excvecprime}
e'\exp{i\omega'}=e_0'\exp{i(\omega_0'+g't)}+\sum_{k=1}^{S} A_k
\exp{i(\omega_{k0}'+g_kt)}
\end{equation}
\begin{equation}\label{excvec} 
e_j\exp{i\omega_j}=B_{0j}\exp{i(\omega_{0j}+g't)}+\sum_{k=1}^{S}
B_{kj} \exp{i(\omega_{kj}'+g_kt)}~~.
\end{equation}
\noindent
Without loss of generality , the constant $\omega_{0}'$ can be set
equal to zero. The positive quantities $A_{k}$, $B_{kj}$, with $k=1,...,S$, 
and $B_0$, are hereafter collectively referred to as 'the amplitudes of
oscillation of the planetary eccentricities'. Also, we assume that a
typical condition over the solutions (\ref{excvecprime}) holds,
namely, that at least for the giant primary one has $e_0' >
\sum_{k=1}^{S} A_k$. Then, at least the giant primary exhibits a
constant, on the average, precession of its eccentricity vector, 
by the frequency $g'$, i.e., one can write $e' = e_{0}' + F$, 
$\omega' = g' t + G$, where the functions $F$ and $G$ depend 
trigonometrically on the angles $\phi'=g't$, $\phi_{j}=g_{j} t$, 
$j=1,...,S$, while their size is of the order of the amplitudes 
$A_k$, $k=1,...,S$.

For each secular frequency we now introduce a pair of action-angle 
variables, i.e. the `dummy' actions $I',I_j$ and the angles 
$\phi'=g't$, $\phi_j=g_j t$, $j=1,2,...,S$. The primary's 
elements are given by $e'=e_0'+F(\phi',\phi_j)$, $\omega'=\phi'
+G(\phi',\phi_j)$. The form of the Hamiltonian is found by the 
following steps:\\
\\
i) Add to the Hamiltonian (\ref{ham4}) the direct terms for $S$ 
planets. The direct term for the j-th planet is:
$$
R_{j,direct}=-\mu_j \left({1\over\Delta_j}
-{\mathbf{r}\cdot\mathbf{r_j}\over r_j^3}\right)
$$
where $\mu_j=m_j/(M+m')$ with $m_j$ equal to the mass of the 
j-th planet, and $r_j=|\mathbf{r}_j|$, $\Delta_j = \vert \mathbf{r} 
- \mathbf{r}_j \vert$, $\mathbf{r}_j$ being the heliocentric 
position vector of the j-th planet. Transcribed to elements, 
$R_{j,direct}$ depends on $(\lambda,\omega,\lambda_j,\omega_j)$. 
Assuming that the j-th planet is far from a mean motion resonance 
with the primary (and hence with the trojan body), we can compute 
$$
<R_{j,direct}>={1
\over 4\pi^2}\int_{0}^{2\pi}\int_{0}^{2\pi}R_{j,direct}d\lambda d\lambda_j~~.
$$ By rotational symmetry, $<R_{j,direct}>$ depends only on the
difference $\omega-\omega_j$, and hence, (see Eq.(\ref{excvec})), only
on the angles $\omega$, $\phi_j$ and $\phi_j$, $j=1,2,...,S$. By
d'Alembert rules, this implies that it is also of first or higher
degree in the eccentricity $e_j$, i.e., it is of first or higher
degree in the amplitudes of oscillation of the planetary
eccentricities.

ii) Consider now the indirect effects of the S planets on the trojan 
body. Again, far from mean motion resonances one has that the primary's 
major semi-axis remains constant. Then, the indirect effects are 
accounted for by rendering the parameters $e',\omega'$ in the 
expression (\ref{ham4}) time-dependent (according to 
Eq.(\ref{excvecprime})) rather than constants. Replace now 
$e'=e_0'+F(\phi',\phi_j)$, $\omega'=\phi'+G(\phi',\phi_j)$ 
(see above), and Taylor-expand,  around $e_0'$ and $\phi'$, 
assuming $F$ and $G$ small quantities. This leads to:
$$
R(\lambda,\omega,x,y,\lambda',\omega',e')=
R(\lambda,\omega,x,y,\lambda',\phi';e_0') + R_2
$$
where $R_2$ is of degree one or higher in the quantities $F,G$, 
and hence, of degree one or higher in the mass parameters $\mu_j$, 
$j=1,\ldots,S$. 

iii) Adding also terms for all dummy actions $I'$, $I_j$, the final 
Hamiltonian is now written as 
\begin{equation}\label{hamsec}
H=-{1\over 2(1+x)^2} + I_3 + g'I'+ \sum_{j=1}^S g_j I_j - 
\mu R(\lambda,\omega,x,y,\lambda',\phi';e_0') 
- \mu R_2 - \sum_{j=1}^S\mu_j{\cal R}_{j}
\end{equation}
where $R_2$ and ${\cal R}_{j}\equiv <R_{j,direct}>$ contain terms 
of first or higher degree in the amplitudes of oscillation of the planetary
eccentricities.  

Two important remarks are in order: i) The function $R$ in Eq. (\ref{hamsec}) 
is formally identical to the function $R$ in (\ref{ham4}), apart from 
replacing $e'$ with $e_{0}'$ and $\omega'$ with $\phi'$. ii) The 
functions $R_2$ and ${\cal R}_j$, $j=1,\ldots S$ are of first 
or higher degree in the amplitudes of oscillation of the planetary 
eccentricity vectors. These remarks turn to be crucial in 
the characterization of the diffusion process along resonances in the 
domain of trojan motion (see below). At any rate, we note that the case 
where mean motion resonances between the planets are present 
necessitates a separate treatment, since then the domain of co-orbital 
motion can be crossed by resonances of the type of the `great inequality' 
(see Robutel and Gabern 2006).

For studying trojan dynamics, it is important to introduce two 
new angles, namely $\tau=\lambda-\lambda'$ and $\beta=\omega-\phi'$. 
The angle $\tau$ is the resonant angle corresponding to the 1:1 
resonance, with value $\tau=\pi/3$ (at the Lagrangian point $L_4$). The 
angle $\beta$, on the other hand, expresses the relative position 
of the pericenter of the trojan body from the pericenter of the 
primary body, since, one has $\beta=\omega-g't+ {\cal O}(\mu_j)$. The two 
angles are formally introduced by defining new canonical variables 
via the generating function:
\begin{equation}\label{gensec}
{\cal S}_1=(\lambda-\lambda')X+\lambda'J_3+(\omega-\phi')J_2 
+ \phi'P'+\sum_{j=1}^S \phi_jP_j  
\end{equation}
yielding the transformation to new angles
\begin{equation}\label{trfsecang}
\tau=\lambda-\lambda',~~~ q_3=\lambda',
~~~\beta=\omega-\phi',~~~\theta'=\phi',~~~
\theta_j=\phi_j,~j=1,...,S
\end{equation}
and new actions $X,J_2,J_3,P',P_j$ connected to the old ones via
\begin{equation}\label{trfsecact}
x=X,~~~I_3=J_3-X,~~~y=J_2,~~~I'=P'-J_{2},~~~I_j=P_j.
\end{equation}
Note that preserving the canonical character of the variables 
requires some modification of the dummy action variables as well. 
In numerical computations, the transformed dummy variables are 
employed in the numerical determination of proper element values 
(see below). Keeping old notation for all variables involved in a 
identity transformation, we rewrite as  $x$,$\lambda'$,$y$,$I_j$,
$\phi'$,$\phi_j$ the variables $X=x$, $q_3=\lambda'$,$J_2=y$,
$P_j=I_j$,$\theta'=\phi'$, $\theta_j=\phi_j$. The Hamiltonian 
 then reads:
\begin{eqnarray}\label{hamsec2}
H&=&-{1\over 2(1+x)^2} -x + J_3
 - g'y + g'P'+ \sum_{j=1}^S g_j I_j\nonumber - \mu R(\tau,\beta,x,y,
\lambda',\phi';e_0')\\
&-& \sum_{j=1}^S\mu_j {\cal R}_j(x,y,\beta,\phi',\phi_1,...,\phi_{s})\nonumber 
- \mu R_2(x,y,\tau,\beta,\phi',\phi_1,...,\phi_s)\nonumber~~.
\end{eqnarray}

\subsection{Forced equilibrium}
The Hamiltonian (\ref{hamsec2}) can be recast under the form:
\begin{equation}\label{hamsec3}
H=<H>+H_1
\end{equation}
where
$$
<H>=-{1\over 2(1+x)^2} -x + J_3 - g'y -\mu <R>(\tau,\beta,x,y;e_0')
$$
and 
$$
H_1=
 g'P'+ \sum_{j=1}^S g_j I_j -\mu \tilde{R}(\tau,\beta,x,y,\lambda',\phi';e_0')
~~~~~~~~~~~~~~~~~~~
~~~~~~~~~~~~~~~~~~~
$$
$$
-\sum_{j=1}^S\mu_j {\cal R}_j(x,y,\beta,\phi',\phi_1,...,\phi_s)
- \mu R_2(x,y,\tau,\beta,\phi',\phi_1,...,\phi_s)
$$
with 
$$
<R>={1\over 2\pi}\int_{0}^{2\pi}R d\lambda',~~~~\tilde{R}=R-<R>~~.
$$ The exact form of $<H>$ up to order two in the eccentricities is
given in the Appendix. In
the ERTBP, we obtain the same form by replacing $\beta$ with
$\omega-\omega'$ and $e_0'$ with the (constant) $e'$, and setting
$g'=0$. Also, it is easy to check that the angle $\phi'$ does not
appear in $<H>$ as a consequence of the D'Alembert rules. The action
$J_3$ is an integral of motion under the Hamiltonian flow of
$<H>$. Thus, the Hamiltonian $<H>$ represents a system of two degrees
of freedom.  We define the forced equilibrium as the solution to the
system of equations
$$
\dot{\tau}={\partial<H>\over\partial x}=0,~~~
\dot{\beta}={\partial<H>\over\partial y}=0,~~~
\dot{x}=-{\partial<H>\over\partial\tau}=0,~~~
\dot{y}=-{\partial<H>\over\partial \beta}=0~~~.
$$
For $L_4$, one finds that
\begin{equation}\label{forced}
(\tau_0,\beta_0,x_0,y_0)= \big( \pi/3,\pi/3,0,\sqrt{1-e_0'^2}-1 \big) +O(g')~~.
\end{equation}
Note that the position of the forced equilibrium, as determined by
(\ref{forced}), is time-independent.  Nevertheless, the point of
Eq. (\ref{forced}) is not an exact equilibrium point under the
complete hamiltonian, including the terms ${\cal R}_j$ and $R_2$. 
Time-dependent corrections to the position of the forced
equilibrium can be determined via a series of canonical
transformations, or by working directly with the equations of motion
of the complete flow of $H$ (see, for example, Milani 1993, or Morais
2001). Also, Erdi (1997) and Morais (2001) considered corrections 
to the definition of the forced equilibrium depending on a parameter 
$l$, which constitutes a quasi-integral of motion (called 'proper
libration'). However, this integral is not precise for the resonant
motions considered below. Thus, we do not take such corrections 
explicitely into account, while corrections depending on a resonant 
form of the quasi-integral can again be found using canonical 
transformations on the full Hamiltonian model $H$.

\subsection{Expansion around the forced equilibrium}
We will now introduce local action-angle variables around the 
point of forced equilibrium. The purpose is to characterize the 
motion by two approximate constants, one of which appears as an 
action variable ($J_s$) on the plane $(x,\tau)$ around the value 
$(x_0,\tau_0)$, while the other appears as an action variable 
($Y_p$) on the plane $(y,\beta)$ around the value $(y_0,\beta_0)$. 
To this end, we first introduce the `shift of center' canonical 
transformation given by:
\begin{equation}\label{poincvar}
v=x-x_0,~~u=\tau-\tau_0,~~Y=-(W^2+V^2)/2,~~\phi=\arctan(V/W) 
\end{equation}
where
$$
V=\sqrt{-2y}\sin\beta-\sqrt{-2y_0}\sin\beta_0,~~~~
W=\sqrt{-2y}\cos\beta-\sqrt{-2y_0}\cos\beta_0~~.
$$
Re-organising terms, the Hamiltonian (\ref{hamsec2}) takes the form:
\begin{eqnarray}\label{hamsec4}
H&=&-{1\over 2(1+v)^2} -v + J_3  - g'Y\nonumber\\ 
&-&\mu\left(
{\cal F}^{(0)}(u,\lambda'-\phi,v,Y;e_0')+
{\cal F}^{(1)}(u,\phi,\lambda',v,Y;e_0')
\right)\nonumber\\
&+& g'P'- \mu {\cal F}^{(2)}(u,\phi,\lambda',v,Y,\phi';e_0')\\
&+& \sum_{j=1}^S g_j I_j - \sum_{j=1}^S\mu_j 
{\cal F}_j(u,\phi,v,Y,\phi,\phi',\phi_j,\omega_{0j},e_0',e_{0j})\nonumber
\end{eqnarray}
where (i) ${\cal F}^{(0)}$ contains terms depending on the angles
$\lambda'$, $\phi$ only through the difference $\lambda'-\phi$, (ii)
${\cal F}^{(1)}$ contains terms dependent only on non-zero powers of
$e_0'$ and independent of $\phi'$, and (iii) ${\cal F}^{(2)}$
contains terms dependent on $\phi'$ and also on non-zero powers
of either $e_0'$ or the oscillation amplitudes of the planetary
eccentricities. The above decomposition of the Hamiltonian allows to
consider various `levels' of perturbation as follows: We call `basic
model' the one of Hamiltonian
\begin{equation}\label{hambasic}
H_b=-{1\over 2(1+v)^2} -v + J_3 - g'Y -\mu{\cal
  F}^{(0)}(u,\lambda'-\phi,v,Y;e_0')~~~.
\end{equation}
The total Hamiltonian takes the form $H=H_b+H_{sec}$, where
$$
H_{sec}= 
-\mu{\cal F}^{(1)}(u,\phi,\lambda',v,Y;e_0')
+ g'P'- \mu {\cal F}^{(2)}(u,\phi,\lambda',v,Y,\phi';e_0')
$$
$$
+ \sum_{j=1}^S g_j I_j - \sum_{j=1}^S\mu_j 
{\cal F}_j(u,\phi,v,Y,\phi,\phi',\phi_j,\omega_{0j},e_0',e_{0j})~~.
$$

The important remark is, again, that the Hamiltonian $H_b$ is
  formally identical in the RMPP and in the ERTBP, with the
  substitution $e_0'\rightarrow e'$ and setting $g'=0$. Furthermore,
the fact that in (\ref{hambasic}) the angles $\lambda',\phi$ appear
only under the combination $\lambda'-\phi$ implies that, in both
  cases, $H_b$ can be reduced to a system of two degrees of
  freedom. The reduction is realized by the canonical transformation:
\begin{equation}\label{gencirc}
{\cal S}_2(u,\lambda',\phi,Y_u,Y_s,Y_p) = u Y_u + (\lambda'-\phi) Y_f
+ \phi Y_p
\end{equation}
yielding
$$
\phi_u={\partial {\cal S}_2 \over\partial Y_u}=u,~~~ \phi_f={\partial
{\cal S}_2\over\partial Y_f}=\lambda'-\phi,~~~ \phi_p={\partial
{\cal S}_2 \over\partial Y_p}=\phi,
$$
and
$$
v={\partial {\cal S}_2 \over\partial u}=Y_u,~~~ J_3={\partial
{\cal S}_2 \over\partial \lambda'}=Y_f,~~~ Y={\partial {\cal S}_2 \over\partial
\phi}=Y_p-Y_f~~.
$$ 
The subscripts `f' and `p' stand for `fast' and `proper' respectively, 
for reasons explained below. As before, we keep the old notation 
for the variables transforming by the identities $\phi_u=u,\phi_p=\phi$, 
and $Y_u=v$. However, it turns to be convenient to retain the new 
notation for the action $Y_f\equiv J_3$.  The Hamiltonian $H_b$ in the 
new canonical variables reads
\begin{equation}\label{hambasic2}
H_b=-{1\over 2(1+v)^2}-v + (1+g')Y_f -g'Y_p
-\mu{\cal F}^{(0)}(u,\phi_f,v,Y_p-Y_f;e_0')~~~.
\end{equation}
Since $\phi$ is ignorable, $Y_p$ is an integral of the Hamiltonian 
(\ref{hambasic2}). The physical importance of $Y_p$ can be understood 
as follows: The action variable $Y$ measures the radial distance from 
the point of forced equilibrium in the plane $(V,W)$, in which the 
forced equilibrium is located at the origin. In a first approximation, 
the quasi-integral of the proper eccentricity can be defined as
\begin{equation}\label{epr}
e_{p,0}=\sqrt{V^2+W^2}=\sqrt{-2Y}~~.
\end{equation}
However, the above definition neglects the fact that $Y$ is subject 
to fast variations due to its dependence on $Y_f$. In fact, by the 
Hamiltonian equations of motion we readily find that $\dot{Y}_f= 
{\cal O}(\mu)$. The time variation of $Y_f$ is associated 
to a fast frequency ${\dot{\phi_f}=1-g}$. In fact, by their 
definition we can see that the variables $(\phi_f,Y_f)$ describe 
epicyclic oscillations of the trojan body, i.e. $\phi_f$ accomplishes 
one cycle every time when the trojan body passes through a local 
pericenter. As shown with numerical examples below, the time 
variations of $Y_f$ become particularly important when one of the 
following two conditions holds: i) $e_0'<\mu$, or ii) the orbit of 
the trojan planet is subject to a low-order resonance. On the other 
hand, $Y_p$ remains an exact 
integral of the Hamiltonian (\ref{hambasic2}) even in the cases 
(i) or (ii), while it is subject only to secular variations in the 
full Hamiltonian model (\ref{hamsec4}). We thus adopt the following 
definition of the proper eccentricity:
\begin{equation}\label{eprnew}
e_p = \sqrt{-2Y_p}~~.
\end{equation}
In the Hamiltonian (\ref{hambasic2}), the integral $Y_p$ (or $e_p$) 
becomes a label of a system of two degrees of freedom corresponding 
to the canonical pairs $(u,v)$ and $(Y_f,\phi_f)$. Collecting terms 
from ${\cal F}^{(0)}$ linear in $Y$, we find:
\begin{equation}\label{freqs}
\omega_f\equiv\dot{\phi_f} = {\partial H_b\over\partial Y_f} = 
1-{27\mu/8}+g'+...,~~~
g\equiv\dot{\phi} = {\partial H_b\over\partial Y_p} = 
{27\mu/8}-g'+...~~~
\end{equation}
We identify $\omega_f$ and $g$ as the short-period and secular 
frequencies, respectively, of the trojan planet. A second averaging 
over the fast angle $\phi_f$ yields the Hamiltonian
\begin{equation}\label{hambasic3}
\overline{H_b}(u,v;Y_f,Y_p,e_0')=-{1\over 2(1+v)^2}-v + (1+g')Y_f -g'Y_p
-\mu\overline{{\cal F}^{(0)}}(u,v,Y_p-Y_f,e_0')~~
\end{equation}
with
$$
\overline{{\cal F}^{(0)}}={1\over 2\pi}\int_{0}^{2\pi}{\cal F}^{(0)}d\phi_f~~.
$$ The Hamiltonian $\overline{H_b}(u,v;Y_f,Y_p,e_0')$ represents a
system of one degree of freedom, all three quantities $Y_f,Y_p,e_0'$
serving now as parameters, i.e. constants of motion under the dynamics
of $\overline{H_b}$.  The Hamiltonian $\overline{H_b}$ describes the
synodic (guiding-center) motions of the trojan planet 
\footnote{The variables $u,v$ proposed here are similar to the variables 
$X,Y$ defined in equations (24) or (25) of Beaug\'{e} and Roig (2001), 
via a so-called `Jupp transformation'. Note, however, the following
difference: Our Hamiltonian $\overline{H_b}$, analogous to Beaug\'{e} 
and Roig's $F_0$, is, nevertheless, labeled also by $e_0'$ (or, simply 
$e'$ in the ERTBP). Thus, it is {\it not} the averaged (over fast angles) 
Hamiltonian of the circular RTBP. From a physical point of view, this 
expresses the possibility to find an integrable approximation to synodic 
motions even when $e'\neq 0$. Accordingly, $Y_p$ in our formalism, which 
is analogous to Beaug\'{e} and Roig's $W$, provides a label for the proper
eccentricity rather than simply the eccentricity of the test particle. 
Again, this expresses the fact that the former, but not the latter, 
is nearly constant even when $e'\neq 0$.}. 
The equilibrium point $(u_0,v_0)$ given by
$$
{\partial \overline{{\cal F}^{(0)}}\over\partial u_0}=
{\partial \overline{{\cal F}^{(0)}}\over\partial v_0}=
0
$$
corresponds to a short-period periodic orbit of the Hamiltonian $H_b$ 
around the forced equilibrium point. We define the action variable 
\begin{equation}\label{synodic}
J_s={1\over 2\pi}\int_C (v-v_0)d(u-u_0)
\end{equation}
where the integration is over a closed invariant curve $C$ around 
$(u_0,v_0)$ (`s' stands for `synodic'). The angular variable $\phi_s$, 
conjugate to $J_s$, evolves in time according to the synodic frequency 
$\omega_s$ given by (see Eq.(\ref{omesyn}) in subsection 2.5 below)
\begin{equation}
\omega_s=\dot{\phi}_s=\sqrt{27\mu\over 4}+... 
\end{equation}

\begin{figure}
\begin{center}
\includegraphics[width=0.85\textwidth,angle=0]{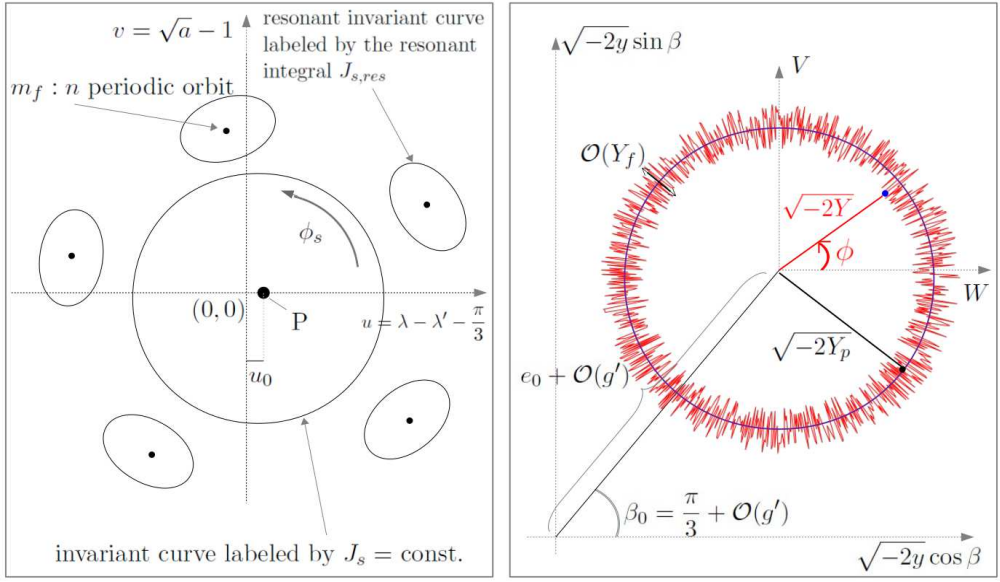}
\end{center}
\caption{{\small Schematic representation of the physical meaning of
    the action-angle variables introduced in subsections 2.2. and 2.3.
    The plane $(u,v)$ corresponds to the `synodic' motion of the
    trojan body. Under the Hamiltonian $H_b$, the phase portrait can
    be represented by a Poincar\'{e} surface of section corresponding,
    e.g., to every time when the angle $\phi_f$ accomplishes a full
    cycle. The left panel shows schematically the form of the
    projection of this section on the plane $(u,v)$. The central point
    P represents a stable fixed point corresponding to the
    short-period periodic orbit around L4. The orbit has frequency
    $\omega_f$, while its amplitude increases monotonically with
    $Y_f$. The forced equilibrium corresponds to $u_0=0$, $Y_f=0$. The
    point P, however, has in general a shift to positive values
    $u_0>0$ for proper eccentricities larger than zero (see subsection
    2.5 below). Far from resonances, the invariant curves around P are
    labeled by a constant action variable $J_s$, and its associated
    angle (phase of the oscillation) $\phi_s$. Resonances, and their
    island chains correspond to rational relations between the frequencies
    $\omega_f$ and $\omega_s$. Within the resonant islands, $J_s$ is
    no longer preserved, but we have, instead, the preservation of a
    resonant integral $J_{s,res}$.  On the other hand, the plane
    $(W,V)$ (right panel) depicts the evolution of the trojan body's
    eccentricity vector under the Hamiltonian $H_b$. The motion of the
    endpoint of the eccentricity vector can be decomposed to a
    circulation around the forced equilibrium, with angular frequency
    $g$, and a fast (of frequency $\omega_f$) `in-and-out' oscillation
    with respect to a circle of radius $e_p$, of amplitude which is of
    order ${\cal O}(Y_f)$. Under $H_b$ alone, the quantities
    $Y_p,J_s$, or $Y_p,J_{s,res}$ are quasi-integrals for all the
    regular orbits. Since $H_b$ is formally identical in the ERTBP and
    in the multi-planet restricted problems, we conclude that the
    basic features of dynamics induced by $H_b$ apply in the same way
    with or without additional planets. Furthermore, all extra terms
    with respect to $H_b$ in the Hamiltonian (\ref{hamsec2}) depend on
    the slow angles $(\phi,\phi')$ and, in the case of the
    multi-planet problem, $\phi_j$, $j=1,\ldots,S$. Thus, all these
    terms can only slowly modulate the dynamics under $H_b$. In the
    case of the ERTBP, the modulation can produce a long-term drift of
    the values of $(Y_p,J_s)$, or $Y_p,J_{s,res}$. This is exemplified
    by numerical simulations in the sequel. The drift can lead to
    large long term variations of the actions, and eventually to an
    escape of the trojan body. A similar drift is produced in the
    multi-planet restricted problem. In the latter case, we have
    additionally that the position of the forced equilibrium
    oscillates quasi-periodically around the origin of the $(W,V)$
    system of axes. The amplitude of oscillation is of order ${\cal
      O}(max(\mu_j e_j))$, while the frequency is of first order in
    the planetary masses.}}
\label{schem}
\end{figure}

Figure (\ref{schem}) summarizes the physical meaning of the 
various action-angle variables introduced so far. We emphasize that 
in numerical computations one always stays with the original (Cartesian) 
co-ordinates of the various bodies. Then, translation of the results to 
action-angle variables and vice-versa is straightforward, passing first 
to Delaunay elements, and then using the transformations (\ref{gensec}, 
\ref{poincvar}, and \ref{gencirc}).

\subsection{Resonances}
The most general form of planar secondary resonances is given by the 
commensurability relation:
\begin{equation}\label{resgen}
m_f\omega_f+m_s\omega_s+m g + m'g'+m_1g_1+\ldots+m_Sg_S=0
\end{equation}
with $m_f,m_s,m,m',m_j$ (with $j=1,\ldots,S$) integers. Of those,
the most important are the resonances of the basic Hamiltonian 
$H_b$. These resonances exist in the complete spectrum of possible 
problems, from the planar circular restricted three body
problem ($S=0$, $g'=0$, $e_0'=0$) up to the complete multi-planet 
problem. They are of the form
$m_f=1$, $m_s<0$, $m=0$, i.e.
\begin{equation}\label{resba}
\omega_f-n\omega_s=0
\end{equation}
with $n=-m_s$.  We briefly refer to a resonance of the form (\ref{resba}) 
as the `1:n' resonance, and to higher order resonances as the $m_f:n$ 
resonances. For $m_f=1$ and $\mu$ in the range $0.001\leq\mu\leq 0.01$, 
$n$ is in the range $4\leq n\leq 12$. In the frequency space  
$(\omega_f,\omega_s,g)$, the relations (\ref{resba}) represent planes 
normal to the plane $(\omega_f,\omega_s)$ which intersect each other 
along the $g$--axis. In the same space, all other resonances with 
$|m|+|m'|+|m_1|+\ldots+|m_S|>0$ intersect transversally one or more 
planes of the main resonances. We refer to such resonances as `transverse'
if $|m_f|+|n|>0$, or `secular' if $|m_f|+|n|=0$. In the numerical
examples below, we use the notation $(m_f,m_s,m)$, for the integers of
the resonant condition (\ref{resgen}) setting $m'=m_1=...=m_s=0$ for
the case of the ERTBP. As argued above, the dynamical role played 
by transverse resonances involving the proper frequency $g$ is quite
similar to the one played by resonances involving the secular
planetary frequencies $g'$ and $g_j$, $j=1,...,S$. Thus, we do not
introduce any further diversification between {\it transverse}
resonances arising in the ERTBP and those due to planetary secular
dynamics. This suggests that most numerical results found below
regarding the chaotic diffusion at resonances in the case of the ERTBP
apply to the transverse resonances of more general models involving
more than one disturbing planets.
\footnote{A comparison between the terminology for resonances employed 
here and in Robutel and Gabern (2006, referring to the case of Jupiter's 
trojan asteroids), is in order. Robutel and Gabern distinguish four 
`families' of resonances. Two of these, however, (Family II and IV) 
involve the frequencies $\nu_{1,2}=2n_{Saturn}-n_{Jupiter}$ 
and $\nu_{2,5}=5n_{Saturn}-2n_{Jupiter}$, which, due to the Great 
Inequality, both play an important role in the dynamics of Jupiter's 
trojan swarm. Also, Family III involves the asteroids' secular 
frequency $s$, thus it applies only to inclined motions. Here 
we consider systems with a rather planar geometry and far from 
mean motion resonances, for which only Family I-type resonances 
of Robutel and Gabern are relevant. Their $\nu$ corresponds to our 
synodic frequency $\omega_s$. Also, they use $n_5$ (the mean motion  
of Jupiter = 1 in our units), while we use $\omega_f=n_5-g=1-g$, 
which, as explained above, is the frequency of epicyclic oscillations 
of the trojan body. Thus, The Family I of Robutel and Gabern corresponds 
to our definition (\ref{resgen}) if we set $m_f=-1$, $m_s\equiv p$, 
$m\equiv q-1$, $m'\equiv q_5$, $m_1\equiv q_6$. We emphasize, 
however, that due to the particular values of the frequencies for 
the Trojan swarm, Family I of Robutel and Gabern has the restrictions 
$m_f=-1$, $m+m'+m_1=0$. On the contrary, such restriction is not 
present in our computations. On the other hand, our 'secular' 
resonances require both $m_s$ and $m_f$ to be equal to zero. Thus, 
from a dynamical point of view, they are qualitatively closer to 
their Family III.} 

\subsection{Elliptic restricted three body-problem}
In the elliptic planar restricted three body problem we have $S=0$, 
$g'=0$, $e'=e_0'=$~const. Then, $\beta\equiv\omega$, and $v\equiv x$. 
the Hamiltonian (\ref{hamsec4}) takes the form:
\begin{equation}\label{hamell}
H_{ell}=-{1\over 2(1+x)^2} -x + Y_f 
-\mu\left(
{\cal F}^{(0)}(u,\phi_f,x,Y_p-Y_f,e')
+{\cal F}^{(1)}(u,\phi_f,\phi,x,Y_p-Y_f,e')
\right)~~.
\end{equation}
The function ${\cal F}^{(1)}$ contains terms of at least first order 
in $e'$. Hence, $\dot{Y_p}={\cal O}(e')$. This implies that $Y_p$ (or 
$e_p$) remains a good quasi-integral for not very high values of the 
primary's eccentricity. On the other hand, a more accurate 
(${\cal O}(e'^2)$) quasi-integral can be computed by a first order 
adiabatic theory. 

The form of the functions ${\cal F}^{(0)}$ and ${\cal F}^{(1)}$, with
an error ${\cal O}(x) \approx {\cal O}(\mu^{1/2})$, is given in the
Appendix. From this we deduce the shift in position, with respect to
L4, of the fixed point of $\overline{{\cal F}^{(0)}}$, corresponding
to the short-period orbit around L4 (Namouni and Murray 2000). 
The shift is given by $u_0=\tau_0-\pi/3$, where $\tau_0$ is the 
solution of ${\partial\overline{{\cal F}^{(0)}}/\partial\tau_0}=0$. 
We find:
\begin{equation}\label{u0}
u_0={29\sqrt{3}\over 24}e_p^2+... 
\end{equation}
where the error is of order $4$ in the eccentricities
$e_p,e'$. Taylor-expanding $\overline{H}_{b,ell}$
(Eq.(\ref{hambasic3}), for the ERTBP) around $u_0$ up to terms of
second degree in $\delta u =u-u_0$, we find (up to terms of first
order in $\mu$ and second order in the eccentricities):
\begin{eqnarray}\label{hbaveel}
\overline{H}_{b,ell} &=&-\frac{1}{2} + Y_f -\mu\left({27\over 8}+... \right) 
\frac{e_{p,0}^2}{2}\\
&~&-{3\over 2}x^2+...
-\mu\left({9\over 8}+{63e'^2\over 16}+{129e_{p,0}^2\over 64}+...\right)
\delta u^2+...
\nonumber
\end{eqnarray}
where $\frac{e_{p,0}^2}{2} = -Y_p + Y_f$.  
Since $Y_f$ is ${\cal O}(\mu)$, up to terms linear in $\mu$ the 
part
\begin{equation}\label{harmsyn}
H_{syn}=-{3\over 2}x^2-\mu\left({9\over 8}+{63e'^2\over 16}
+{129e_{p}^2\over 64}+...\right)\delta u^2
\end{equation}
defines a harmonic oscillator for the synodic degree of freedom. 
The corresponding synodic frequency is
\begin{equation}\label{omesyn}
\omega_s=\sqrt{
6\mu\left({9\over 8}+{63e'^2\over 16}+{129e_p^2\over 64}+...\right)
}~~.
\end{equation}
On the other hand, the secular frequency is given by
$g={\partial\overline{H}_{b,ell}/\partial Y_p}$. Assuming a harmonic 
solution $\delta u =\delta u_0\cos(\omega_s t+\phi_{0s})$, and 
averaging over the synodic period $<\delta u^2>=\delta u_0^2/2$, 
we find
\begin{equation}\label{gell}
g= \mu\left({27\over 8} + {129\over 64}\delta u_0^2 +...\right)~~.
\end{equation}

Eq.(\ref{gell}) applies for orbits in the neighborhood of the 
short period orbit. A resonant periodic orbit $m_f:n$ bifurcates 
from the short-period orbit at $\delta u=0$ provided that
\begin{equation}\label{bifres}
m_f\left( 1- {27\mu \over 8} +...\right)=
n\sqrt{
6\mu\left({9\over 8}+{63e'^2\over 16}+{129e_p^2\over 64}+...\right)
}~~.
\end{equation}
Such orbits appear in pairs, one stable and one unstable, and it is
well known that they form `bridges' connecting a short period with a
long period orbit (Deprit and Rabe, 1968). Under the full
Hamiltonian dynamics of $H_{ell}$, the bifurcation generates a
2D-torus, which is the product of the above orbit times a circle on
the plane $W,V$ with frequency $g = 27\mu/8$. 

Under the Hamiltonian flow of $H_b$, a $m_f:n$ resonant periodic orbit 
forms $n$ fixed points on a surface of section $(u,x)$, for 
$mod(\phi_f,2\pi)=const$. Around the fixed points of a stable resonant 
periodic orbit there are formed islands of stability (see figure 
\ref{schem}, schematic), surrounded by separatrix-like thin chaotic 
layers formed around the unstable fixed points. Beyond the 
bifurcation point, as $e_p$ increases, the fixed points move 
outwards, i.e., at larger distances from the central fixed point 
$(x,u)=(0,u_0)$,  while the resonant islands of stability grow in 
size. The growth is faster for lower-order resonant periodic orbits 
(i.e. for smaller $n$). This growth, however, stops when the islands 
of stability enter in the main chaotic sea around the tadpole domain 
of stability. Numerically computed examples of this behavior are given 
in section 3 below. 

\subsection{Modulational diffusion}
The resonant periodic orbits arising under the flow of $H_b$ correspond 
to resonant 2D tube tori under the flow of the full model $H_{ell}$. 
Respectively, the resonant fixed points correspond to one-dimensional 
tori on a surface of section $mod(\phi_f,2\pi)=const$.  Projected 
on the plane $(u,x)$, these tori appear as thick curves (see 
Fig.\ref{diffmodel} below). In the same projection, the islands of 
stability and their delimiting separatrix-like chaotic layers are 
observed to undergo `pulsations', i.e. some periodic shift in the 
plane $(u,x)$ modulated at the frequency $g$ and its multiples. 
Such pulsation is induced by the presence (in $H_{ell}$ but not in $H_b$) 
of terms trigonometric in the angle $\phi$ and its multiples.  

The modulation of all resonant motions by slow trigonometric terms
results in a long-term chaotic diffusion taking place in the space of
the action variables $(J_s,Y_p)$. In fact, based on the pulsation
width of the separatrices, we encounter the following two diffusion
regimes:

i) {\it Non-overlaping resonances}: for small pulsation widths, the
separatrices of one resonance do not enter to the pulsation domain of
the separatrices of nearby resonances. In such cases the rate of the
chaotic diffusion is quite small, and the diffusion becomes
practically undetectable.  Also, the geometry of resonances in the
action space is closer to the paradigm of Arnold diffusion (Arnold 1964). 
An example of chaotic orbit in such regime is given in
section 4 below.

ii) {\it Partially-overlaping resonances}: for large pulsation widths, 
the pulsation domains of more than one separatrices of nearby resonances
partially overlap. In this case the rate of chaotic diffusion 
increases dramatically. As shown in section 4, the chaotic orbits in 
the most prominent chaotic layers exhibit a diffusion timescale of the 
order or 1Myr.  The diffusion leads to final escapes from the resonant 
domain and from the overall tadpole domain. However, there is also a weakly
chaotic population exhibiting long times of stickiness, with a
power-law distribution of the stickiness times characteristic of
long-term correlated chaotic motions (see Meiss 1992, p. 843, and 
references therein, also Ding et al. 1990, Cheng et al. 1992). 
At any rate, in most cases we find that the overlapping of 
resonances is not complete, as is, for example, the case of 
resonant multiplets for Jupiter's trojan asteroids (see Robutel 
and Gabern 2006). As a result, the overall diffusion process in 
our experiments is closer to the paradigm of {\it modulational} 
diffusion (Chirikov et al. 1985). Finally, there are regular resonant 
orbits that never escape the system.
\begin{figure}
\begin{center}
\includegraphics[width=0.90\textwidth,angle=0]{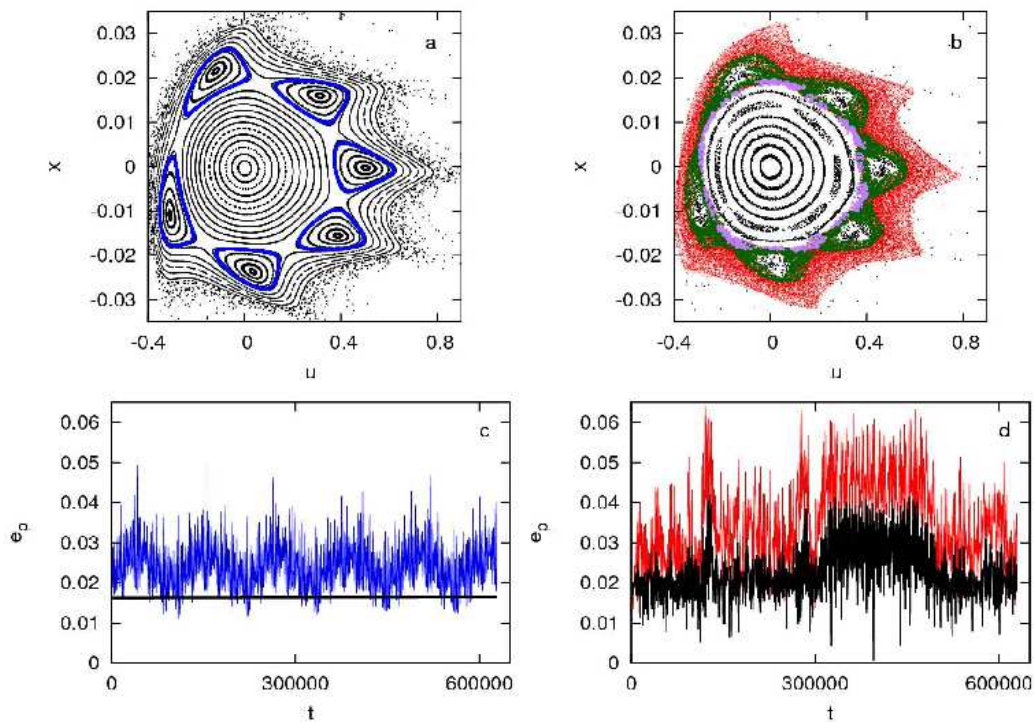}
\end{center}
\caption{(a) - Phase portrait (pericenter surface of section) in the
  variables $u$, $x$, when $\mu=0.0041$, $e_p=0.01675$, $e'=0$
  (circular case). A orbit moving in the thin separatrix layer of the
  $1:6$ resonance is shown in blue (with initial condition $x=0$,
  $u=0.376$).  (b) - Same as in (a) but now in the elliptic
  case $e'=0.02$. The chaotic orbit moves in the separatrix layer of
  the $1:6$ resonance up to a time $10^4$ (green), but later it
  expands towards the chaotic layers of other adjacent resonances
  (red). (c) Time evolution of $e_p$ (Eq. (\ref{eprnew}), black curve)
  and $e_{p,0}$ (Eq. (\ref{epr}), blue curve), for the blue orbit of
  (a). (d) - Time evolution of $e_p$ (black) and $e_{p,0}$ (red) for
  the coloured chaotic orbit of (b).}
\label{diffmodel}
\end{figure}

Figure \ref{diffmodel} gives a typical example of the modulational
diffusion regime. The panel (a) shows a `pericentric' surface of
section $(u,x)$ (precise definition given in section 3 below) which
depicts the structure of the phase space in the circular model
($e'=0$ for the dynamics under the Hamiltonian $H_b$), 
when $\mu=0.0041$, $e_p=0.01675$. For these
parameters, the phase portrait is dominated by the islands of the 1:6
resonance. The separatrix-like chaotic layers surrounding the
resonance are very thin, while the resonant islands are delimited by
both inner and outer librational KAM curves. Thus, all orbits inside this
resonance cannot communicate with orbits of nearby resonances embedded
either in the remaining part of the stability domain or in the chaotic
sea surrounding the stability domain.  An orbit near the
separatrix layer of the $1:6$ resonance is shown in blue in
Fig.\ref{diffmodel}a.  Note that $e_p=\sqrt{-2Y_p}$ is an exact
integral of motion of the flow under $H_b$, as confirmed by a
numerical computation of $e_p$ (Fig.\ref{diffmodel}c, black curve). On
the contrary, the distance $(W^2+V^2)^{1/2}$ from the forced
equilibrium which coincides with the osculating value of the
eccentricity $e$ (Fig.\ref{diffmodel}c, blue curve) undergoes
substantial short-period oscillations (of order $O(\mu)$). Thus, $e_p$
as defined via Eq. (\ref{eprnew}) is a much better measure of the
proper eccentricity than the usual definition $(W^2+V^2)^{1/2}$. 

Figure \ref{diffmodel}b, now, shows the projection on the plane 
$(u,x)$ of a 4D pericentric surface of section in the elliptic case,
 when $e'=0.02$ (black 
points), on which we superpose in colors the points of {\it one} chaotic 
orbit undergoing modulated diffusion. The 1:6 resonance is still clearly 
visible on the $(u,x)$ projection, giving rise to six islands, 
one of which intersects the line $x=0$ at values around $u\approx 0.4$. 
Another smaller 6-island chain, intersecting the line $x=0$ at about 
$u\approx 0.3$ is also distinguishable. As shown in section 4, the 
latter corresponds to the transverse resonance $(1,-6,-1)$, whose extent, 
however is limited and produces no substantial overlapping with other 
low order resonances. On the other hand, the pulsation of the separatrix 
of the 1:6 resonance does results in a substantial overlapping of this 
resonance with other outer resonances surrounding the origin. As a 
result, an orbit started in the separatrix layer of the 1:6 resonance 
later communicates with the separatrix layers of the outer resonances. 
In the example of Fig.\ref{diffmodel}b, the orbit with initial conditions 
indicated in the caption remains confined in the neighborhood of the 
1:6 resonance up to a time $\sim 2\pi\times 10^4$ (green points), while 
at later times (up to $\sim 2\pi\times 10^5$) the same orbit expands 
to embed several higher order resonances of the form $m_f:n$ as well 
as some transverse resonances of the elliptic problem. A careful 
inspection shows that the orbit undergoes several outward and 
inward motions in the whole domain from the 1:6 resonance up to 
the outer resonances, while the orbit eventually escapes from the 
system at still larger times (of order $10^6$). The various outward 
or inward transitions are abrupt, and they are marked by corresponding 
transitions in the value of $e_p$, which is now only an approximate 
adiabatic invariant. Such transitions are shown in Fig.\ref{diffmodel}d 
(black curve). Here, an overall comparison with the time evolution 
of the quantity $(W^2+V^2)^{1/2}$ (red curve), shows that the definition 
of $e_p$ via the action variable $Y_p$ still yields a useful measure of 
the proper eccentricity, while $(W^2+V^2)^{1/2}$ presents wild variations 
even in short timescales. In fact, the time behavior of $e_p$ presents 
jumps at all outward or inward transitions of the corresponding orbit of 
Fig.\ref{diffmodel}b. A further analysis of how the diffusion progresses 
in the space of action variables $(J_s,Y_p)$ is given in section 4.

\section{Stability Maps and resonant structures}

The present section contains a survey of the resonant structures 
appearing in the space of proper elements $(J_s,Y_p)$ for an ensemble 
of parameter values $(\mu,e')$ representative of cases in which a 
resonance of the form $1:n$, with $n=5,\ldots,12$, and its associated 
multiplets, dominates in phase space. A survey of phase portraits 
allows to define a proper set of initial conditions choosen in a grid 
providing a convenient representation of the action variables 
$(J_s,Y_p)$. An atlas of stability maps are then computed by 
means of a suitable chaotic indicator.  

\subsection{Phase portraits and choice of initial conditions}
\label{sec:Section3}
Phase portraits are visualized by means of appropriate Poincar\'{e} 
surfaces of section. In numerical experiments, it turns practical to 
consider {\it apsidal} sections, in which the orbits pass through  
consecutive local pericentric or apocentric positions. Here we adopt 
the pericenter crossing condition, $\dot{r} = 0$ and $\ddot{r} > 0$, 
where $\dot{r}$ is the radial velocity in the heliocentric frame. 

The pericentric surface of section is two-dimensional if $e'=0$, and
four-dimensional if $e'>0$. In the former case, the term
$\mathcal{F}^{(1)}$ in the Hamiltonian (\ref{hamell}) becomes equal to
zero by identity.  The exact invariance of $Y_p$ is equivalent to the
exact invariance of the Jacobi constant $C_J$ in the barycentric
rotating frame. In practice, it is more convenient to construct
surfaces of section of constant values of $C_J$ rather than $Y_p$.
Yet, we label these surfaces of section using a corresponding value of
$e_p = \sqrt{-2Y_p}$. This correspondance is established as follows:
to a given value of $Y_p$ corresponds a short-period orbit crossing
the surface of section at a fixed point with coordinate $u_0$ given by
Eq.(\ref{u0}) with $e_p=\sqrt{-2Y_p}$ (Eq.(\ref{eprnew})). Noticing
that, for $e'=0$ the angles $\phi$ and $\omega$ coincide,
i.e. $\phi_f=\lambda'-\omega$, the remaining initial conditions of the
fixed point are given by
\begin{equation}\label{init}
x_0 = 0, \qquad \phi_{f,0} = \lambda'-\omega_0 = -u_0 - 
\frac{\pi}{3}, \qquad Y_f = 0~~.
\end{equation}
The condition on $\phi_f$ is the pericenter crossing condition.
Setting the Delaunay action $y_0$ as $y=Y_p-Y_f=Y_p$, and the angle
$\lambda_0=\lambda'+\pi/3+u_0$, with $\lambda'=0$ at $t=0$, one then
has all four values of the Delaunay variables
$(\lambda_0,\omega_0,x_0,y_0)$, whereby cartesian position and
velocity vectors can be computed. This allows to compute the Jacobi
constant $C_{J0}$ corresponding to the short-period orbit of given
$e_p$. We refer to the whole surface of section with $C_J=C_{J0}$ as
the section corresponding to a `given value of $e_p$' (referred to as
'the proper eccentricity'), although, for fixed $C_J$, $e_p$ actually
changes somewhat as we move on the section away from the point
$(u_0,x_0)$. Now, for any other point $(u,x)$ on the surface of
section, the pericentric condition yields $\omega=u+\pi/3$, while $y$
(and hence the precise value of $Y_p$) can be computed by solving
numerically the Jacobi-constant equation $C_J=C_{J0}$.

We produce surface of section plots taking 35 equispaced initial 
pericentric conditions along a fixed line of the form $x=B(u-u_0)$, 
up to $u = 1.0$, and solving always the equation $C_J=C_{J0}$. 
The inclination $B$ is determined according to a rule explained below.  
For each initial condition, we integrate the orbits and collect 1000 
successive points on the surface of section, plotted in the plane 
$(u,x)$.

\begin{figure}
\vspace{0cm}
\includegraphics[width=0.90\textwidth,angle=0]{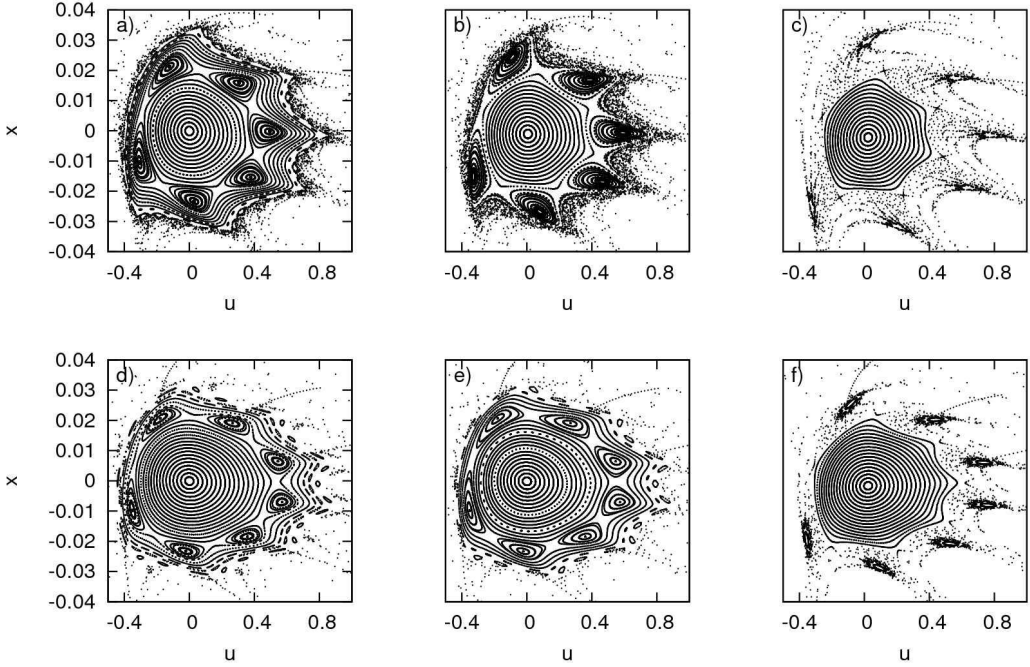}
\caption{Surface of section in the circular case ($e'=0$) for
  $\mu\,=\,0.0041$ (upper plots) and $\mu\,=\,0.0031$ (lower
  plots). The values of $e_p$ are, in each case: $e_p\,=\,0.0001$ (a),
  $e_p\,=\,0.06$ (b), $e_p\,=\,0.1$ (c), $e_p\,=\,0.0001$ (d),
  $e_p\,=\,0.05$ (e) and $e_p\,=\,0.1$ (f).}
\label{pss67}
\end{figure}
Repeating this process for different values of $\mu$, from $0.001$ 
to $0.01$, with an interval of $\Delta \mu = 0.001$, and with 
$e_p=0$, allows to explore the whole range of values containing 
the resonances of the form 1:n, with $4\leq n\leq 12$. 
In particular, between $\mu = 0.0012$ and $\mu = 0.0056$ we find the 
resonances for $n=12,11,...,5$. Higher order resonances of the form 
2:(2n+1), 3:(3n+1) or 3:(3n+2), etc., are 
also distinguishable by simple visual inspection. 

An example is shown in Figure \ref{pss67}. The plots for $\mu=0.0031$ 
(lower row) and $\mu = 0.0041$ (upper row) correspond to the pericentric 
Poincar\'e surfaces of section in two cases where the resonances 1:7 
and 1:6, respectively, are conspicuous in phase space. We note that 
one of the fixed points of the stable periodic orbit corresponding 
to the 1:6 resonance lies on the horizontal line $x=0$. This is so 
for all even resonances (i.e. $1:n$ with $n$ even). On the other hand,  
for odd resonances ($1:n$ with $n$ odd) all stable fixed points 
lie on lines of the form $x=B(u-u_0)$ with $B\neq 0$. In our 
numerical study of FLI maps, we use the following slopes computed 
numerically by inspecting surfaces of sections similar to 
Fig.\ref{pss67}: 
\begin{center}
\begin{tabular}{l c r}\label{beta}
Resonance & $\mu$ & slope $B$\\
\hline
$1$:$11$ & 0.0014 & 0.04 \\
$1$:$9$ & 0.0021 & 0.025 \\
$1$:$7$ & 0.0031 & 0.015 \\
$1$:$5$ & 0.0056 & 0.03 \\
\end{tabular}
\end{center}

A resonant periodic orbit 1:n bifurcates from the central short-period 
orbit at pairs of values $(\mu,e_p)$ satisfying Eq.(\ref{bifres}). 
As shown in Fig.\ref{pss67}, for fixed $\mu$, the resonant orbits 
move outwards as $e_p$ increases, while their corresponding island 
chains grow in size. The three panels in each row of Fig.\ref{pss67} 
correspond to three different values of $e_p$ (see caption), in increasing 
value from left to right. For small values of $e_p$, the stability islands
in both cases are surrounded by invariant tori. The stability domain
around $(u_0,x_0)$ extends from $u \simeq -0.4$ to $u \simeq 0.8$, for
$x=0$. Some small higher order resonances are visible at the border of
the stability domain. However, as $e_p$ increases, the resonant
islands grow in size, while most of their surrounding invariant tori
are destroyed. For a critical value of $e_p$, the last KAM torus
surrounding the resonant island chain is destroyed. We find that this
value satisfies $e_{p,crit}<0.1$ in all studied cases. For
$e_p>e_{p,crit}$, the resonant islands are surrounded by the outer chaotic
sea, which penetrates the stability domain closer and closer to the
center. Thus, for $e_p=0.08$ the right boundary of the stability 
domain shrinks to $u=0.4$ or less.

Similar phenomena appear if $e_p$ is kept fixed while varying $\mu$.
As $\mu$ increases beyond the bifurcation value, the stability islands
of the resonance move outwards and increase in size. Reaching a
certain critical value of $\mu$, the last invariant torus at the
border of stability surrounding the islands is destroyed. This
mechanism also shrinks the stability region, although by abrupt
steps. On the other hand, different values of $\mu$ give rise to
different resonances.  Thus, the size of the domain of stability
undergoes abrupt variations connected to the bifurcations of new
resonances (see Erdi et al. 2007 for a quantitative
study of this effect in the case $e_p=0$).

\subsection{FLI stability maps}
When $e'>0$ the pericentric surface of section becomes 4-dimensional. 
Then, projection effects on the plane $(u,x)$ render unclear a detailed 
visualization of the resonant structures (cf. Fig.\ref{diffmodel}b). 
Nevertheless, a convenient visualization is possible in the space of the 
actions $(J_s,Y_p)$. In practice, we demonstrate all results in a space 
of proper elements $\Delta u$ (libration angle) and $e_p$ (proper 
eccentricity), which are in one to one relation with the action variables 
$(J_s,Y_p)$. The quantity $\Delta u$ is defined as follows: for given 
$e_p$, we first determine $u_0$ via Eq.(\ref{u0}). Then, we consider all 
invariant curves around the equilbrium point $(u=u_0,x=0)$ of the one degree 
of freedom Hamiltonian $\overline{H}_b$ (Eq.(\ref{hambasic3})), as well as 
a line of initial conditions $x=B(u-u_0)$, where, in all examples below, 
$B=0$ for even resonances, or as indicated in the table for odd 
resonances. 
We call $u_p$ the point where the invariant curve corresponding to the action 
value $J_s$ intersects the above line of initial conditions. Finally, we set 
$\Delta u=u_p-u_0$. Using the harmonic oscillator approximation of 
Eq.(\ref{harmsyn}), the action $J_s$ can be approximated as $J_s=E_s/\omega_s$, 
where $\omega_s$ is given by Eq.(\ref{omesyn}), while $E_s$ is the oscillator 
energy $E_s=-H_{syn}$ found by substituting the initial conditions to 
Eq.(\ref{harmsyn}). Then, up to quadratic terms in $\Delta u$, one has 
\begin{equation}\label{jsdu}
J_s = \frac{3B^2/2 + \mu \left( 9/8 + 63e'^2/16 + 129 e_p^2/64 \right)}
{\left[ 6\mu \left( 9/8 + 63 e'^2/16 + 129 e_p^2/64 \right) \right]^{1/2}} 
\Delta u^2 + {\cal O}(\Delta u^4)
\end{equation}
Note that for odd resonances, $\Delta u$ is {\it not} equal to the 
half-width $D_p$ of the oscillation of the variable $u$ along the invariant 
curve of $\overline{H}_b$ corresponding to the action variable $J_s$, which 
is used as a standard definition of the proper libration angle. Instead, 
locating the point where the ellipse defined by $E_s=-H_{syn}$ 
intersects the axis $x=0$, we find
\begin{equation}\label{dudp}
D_p =  \left[ \frac{3B^2/2 + \mu \left( 9/8 + 63e'^2/16 
+ 129 e_p^2/64 \right)}{\mu \left( 9/8 + 63 e'^2/16 
+ 129 e_p^2/64 \right)}  \right]^{1/2} \Delta u 
+ {\cal O}(\Delta u^2)
\end{equation}

In numerical simulations, after fixing $\mu$ and $e'$, we chose a 
$400\times 400$ grid of initial conditions in the square 
$0\leq\Delta u\leq 1$, $0\leq e_p\leq 0.1$, setting also 
$x=B\Delta u$ and $y=Y_p+Y_f$ with $Y_f=0$, $\phi_f=-\pi/3$, 
$\phi=\pi/3$. This completely specifies all Delaunay variables 
for one orbit, and hence its initial cartesian position and 
velocity vectors as well as the value of the dummy action $I_3=0$ 
in the Hamiltonian (\ref{ham4}). Finally, we express (\ref{ham4}) 
in the original Cartesian form
\begin{equation}\label{hamnum}
E=H\equiv {p^2\over 2} +I_3 -{1\over r} - \mu \left({1\over\Delta}
-{\mathbf{r}\cdot\mathbf{r}'\over r'^3}-{1\over r}\right)~~
\end{equation}
and keep track of the constancy of the numerical value of the energy 
$E$ as a probe of the accuracy of numerical integrations.  

\begin{figure}
\hspace{-1.5cm}
\includegraphics[width=0.4\textwidth,angle=270]{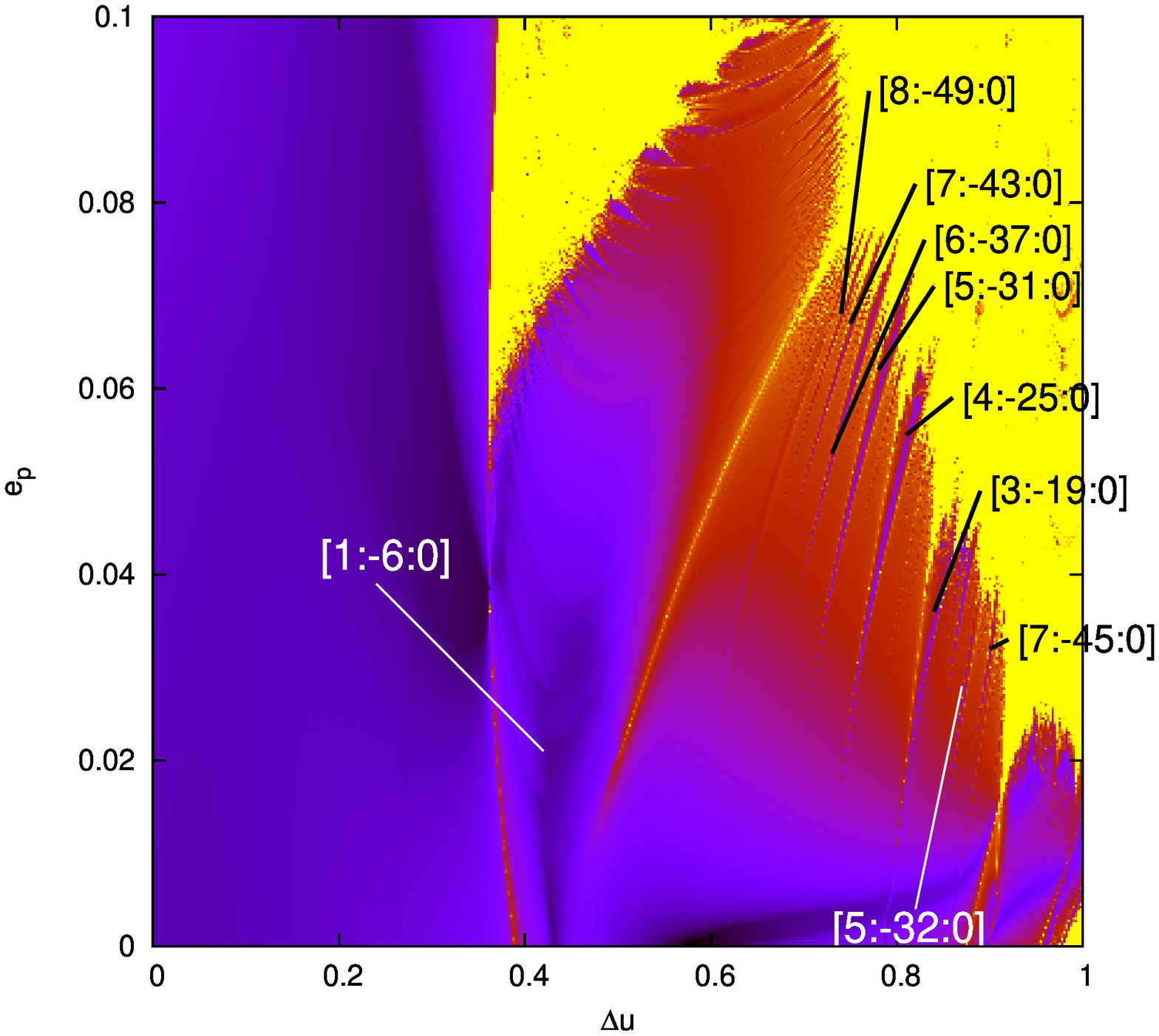}
\includegraphics[width=0.4\textwidth,angle=270]{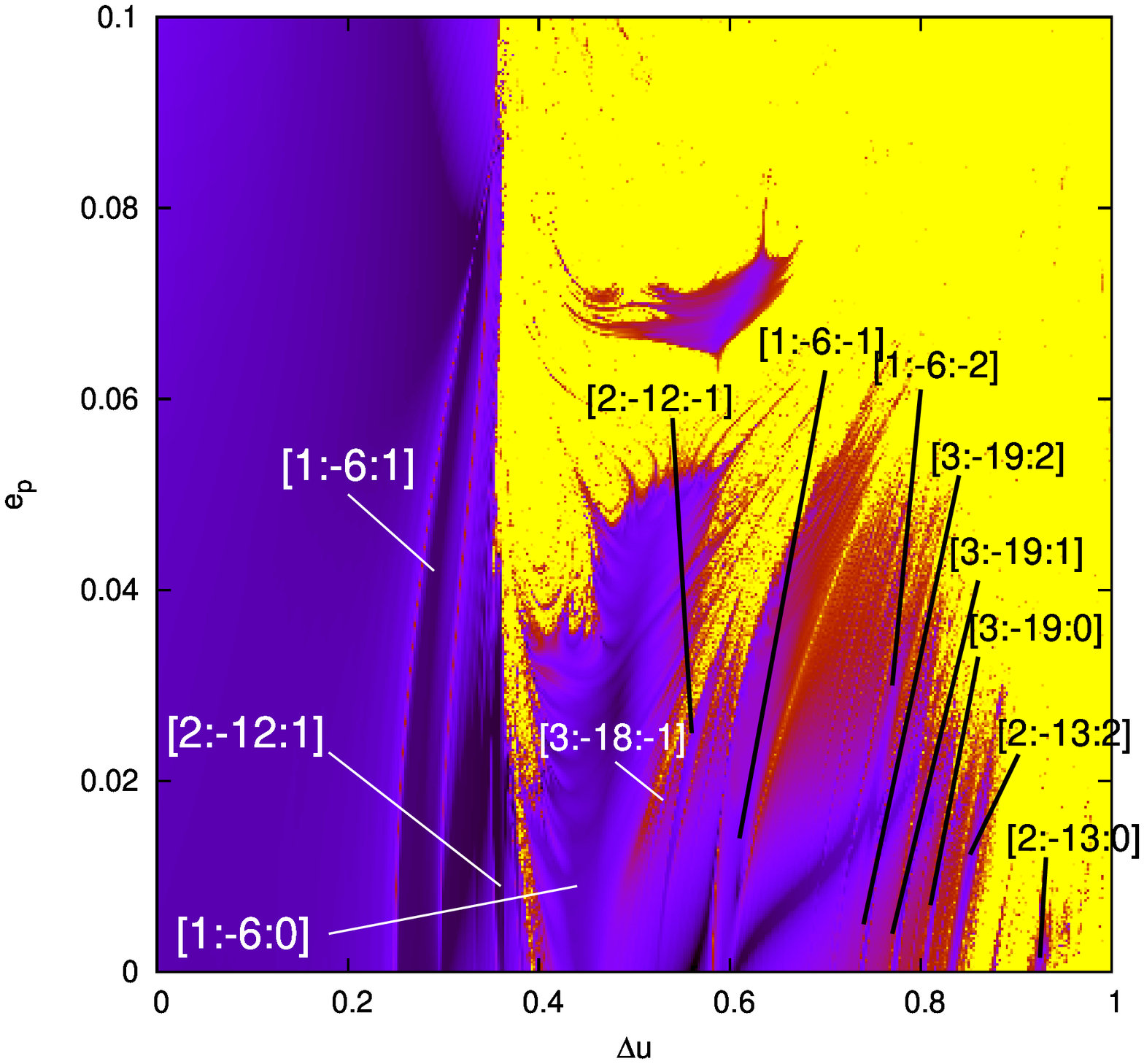}
\begin{center}
\includegraphics[width=0.4\textwidth,angle=270]{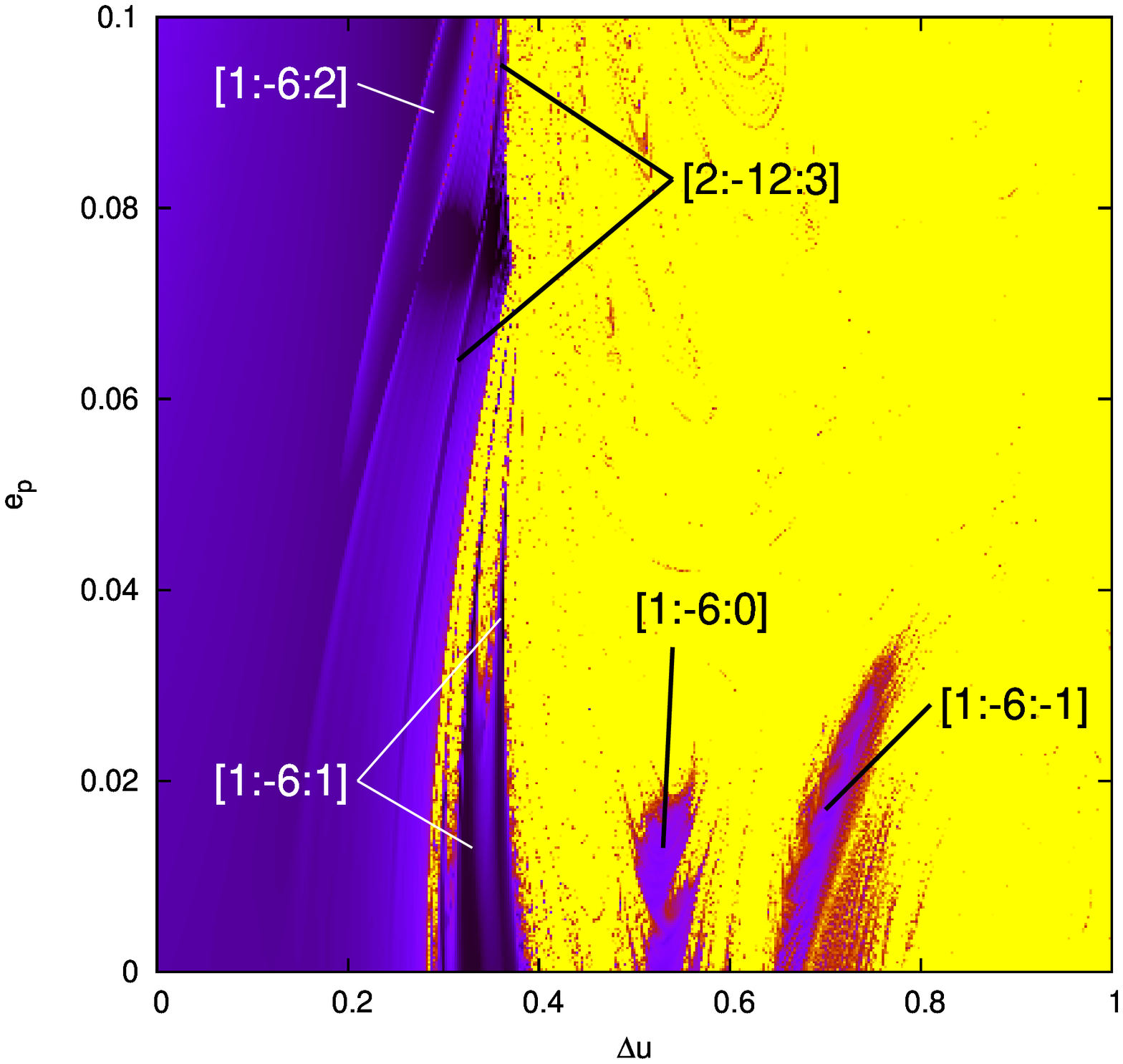}\\
\includegraphics[width=0.8\textwidth]{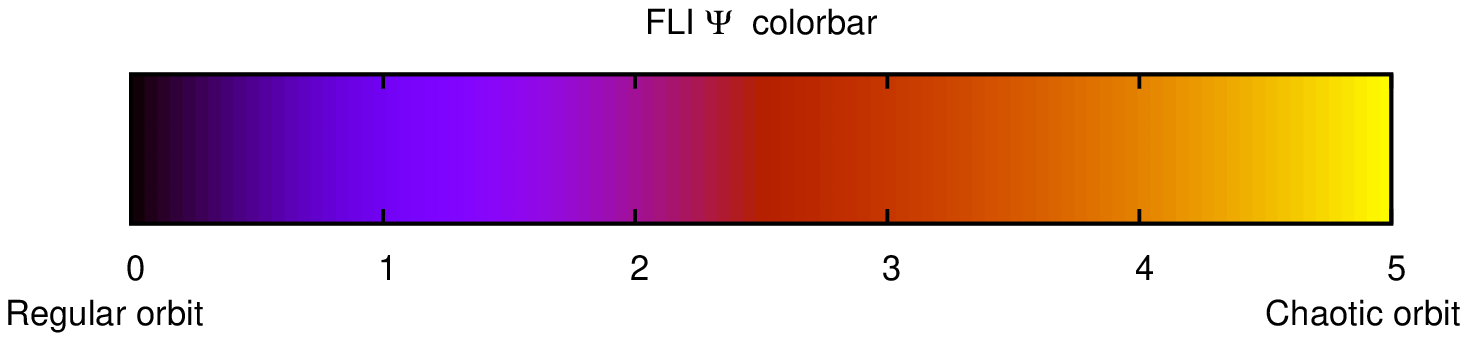}
\end{center}
\caption{FLI maps with details of the resonances, for $e^{\prime}\,=\,0$ 
(left upper panel), $e^{\prime}\,=\,0.02$ (rigth upper panel) and 
$e^{\prime}\,=\,0.06$ (lower panel).}
\label{freqanal}
\end{figure}
Stability maps are computed over the above grid of initial conditions 
by means of color-scaled plots of the value of a suitable chaotic indicator. 
Here we employ the Fast Lyapunov Indicator (FLI, see Froeschl\'{e} et al. 
2000) given by 
\begin{displaymath}
\Psi(t) = \sup_{t} \, \log_{10}( \| {\mathbf \xi} \|) \,, 
\end{displaymath}
where ${\mathbf \xi}$ is the variational (deviation) vector computed 
by solving the variational equations of motion along with the orbital 
equations of motion and $\|\cdot\|$ denotes the $L_2$ norm. For computing
FLIs we implement a $7$th-$8$th order Runge-Kutta method, with a 
timestep equal to $1/300$ of the period of the primary ($=2\pi$ in our 
units). In order to avoid numerical overflows, we stop integrating orbits 
that have clearly reached escape. The latter are identified by a sudden 
jump of the energy error to levels beyond $10^{-4}$, while for non-escaping 
orbits the energy error at $10^3$ periods is less than $10^{-9}$. Also, 
we stop integrating the variational equations of orbits reaching FLI 
values larger than $50$. 

Figure \ref{freqanal} shows an example of computed stability maps for 
$\mu=0.0041$ and three values of $e'$, namely $e'=0$, $e'=0.02$, and 
$e'=0.06$. In each plot, the value of $\Psi$ is given in a color scale 
for all $400\times 400$ grid initial conditions in the plane of proper 
elements $(\Delta u,e_p)$. The color scale was set in the range 
$0\leq\Psi\leq 5$. Regular orbits correspond to darker colors (black) 
representing low values of $\Psi$, while the most chaotic orbits correspond 
to light colors (yellow). Orbits with $\Psi>5$ are shown also in yellow.  

A more detailed resonance identification was made by means of frequency 
analysis (Laskar 1990). The most conspicuous resonances are explicitly 
indicated in all three plots. For $e'=0$ (top left panel), the resonance 
1:6 dominates the stability map. Besides, several resonances of 
the type $(m,-6m+1,0)$ produce strips penetrating the stability domain. 
In agreement with what was shown in the surface of section plots of 
Fig.\ref{pss67}, the width of the 1:6 resonance increases, 
initially, as $e_p$ increases from zero up to a value $e_p\sim0.06$. 
Also, the chaotic separatrix-like layers around the resonance remain 
thin. However, for $e_p>0.06$ the resonance is detached from the main 
stability domain. Then, its corresponding islands of stability are 
embedded in a chaotic sea corresponding to orbits with a fast escape. 
For still larger $e_p$ (around 0.1) the central periodic orbit 
becomes unstable and the corresponding islands disappear. In general, 
this, as well as all higher order resonances move outwards (towards 
higher values of $\Delta u$) as $e_p$ increases. Thus, all resonant 
strips have a small positive slope in Fig.\ref{freqanal}. 

On the other hand, increasing the value of $e'$ causes new {\it
  transverse} resonances to appear. For $e'=0.02$ (top right panel),
resonances of the form $(n,-6n,m)$, or $(n,-6n-1,m)$, with $n=1,2,3$
and $m=-2$ up to $m=2$ are distinguishable. We note that in general
the angle formed between the transverse resonances ($m\neq 0$) and the
secondary resonances of the circular problem ($m=0$) is small (of
order $g/\omega_s$), where $g$ and $\omega_s$ are the respective
secular and synodic frequencies. This small transversality implies
that the intersection point of most mutually transverse resonances
lies in the chaotic zone, i.e. far from the domain of stability. In
particular, the transverse resonances with $m>0$ have no intersection
with the main resonance 1:6 inside the stability domain. In
fact, these transverse resonances penetrate the domain of stability in
isolated, single-resonance strips, along which the orbits undergo
weakly chaotic diffusion that bears many features of Arnold diffusion 
as also shown in Robutel and Gabern (2006). On the other hand, 
the resonances that are beyond the border of the inner stability 
domain form multiplets, in which it has already been argued that 
the chaotic diffusion has features of modulational diffusion. 
This gives a mechanism of efficient escape for chaotic orbits 
(see section 4).

For still higher values of $e'$ (Fig.\ref{freqanal}, down panel for
$e'=0.06$), new transversal resonances appear. In fact, as $e'$
increases all transverse resonances move in the direction from top
left to bottom right, until reaching large values of $\Delta u$, after
which they enter into the main chaotic sea surrounding the domain of
stability. Then, they become less significant.

\subsection{Survey of resonances}

\begin{figure}
\vspace{0cm}
\centering
\includegraphics[width=1.0\textwidth]{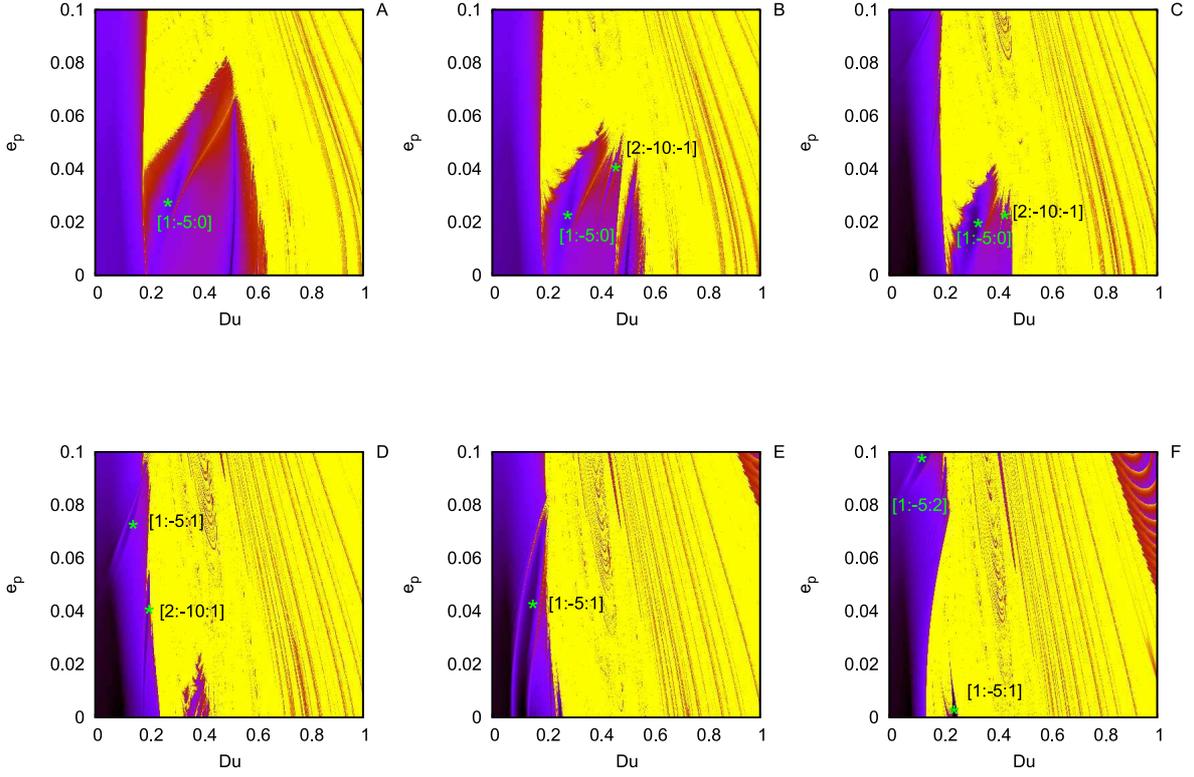}
\vspace{0cm}
\caption{FLI maps for the resonance 1:5, $\mu=0.0056$, 
for the values $e^{\prime}=0$ (A), $e^{\prime}=0.02$ (B), 
$e^{\prime}=0.04$ (C), $e^{\prime}=0.06$ (D), $e^{\prime}=0.08$ 
(E) and $e^{\prime}=0.1$ (F).} 
\label{fli1to5}
\end{figure}

\begin{figure}
\vspace{0cm}
\centering
\includegraphics[width=1.0\textwidth]{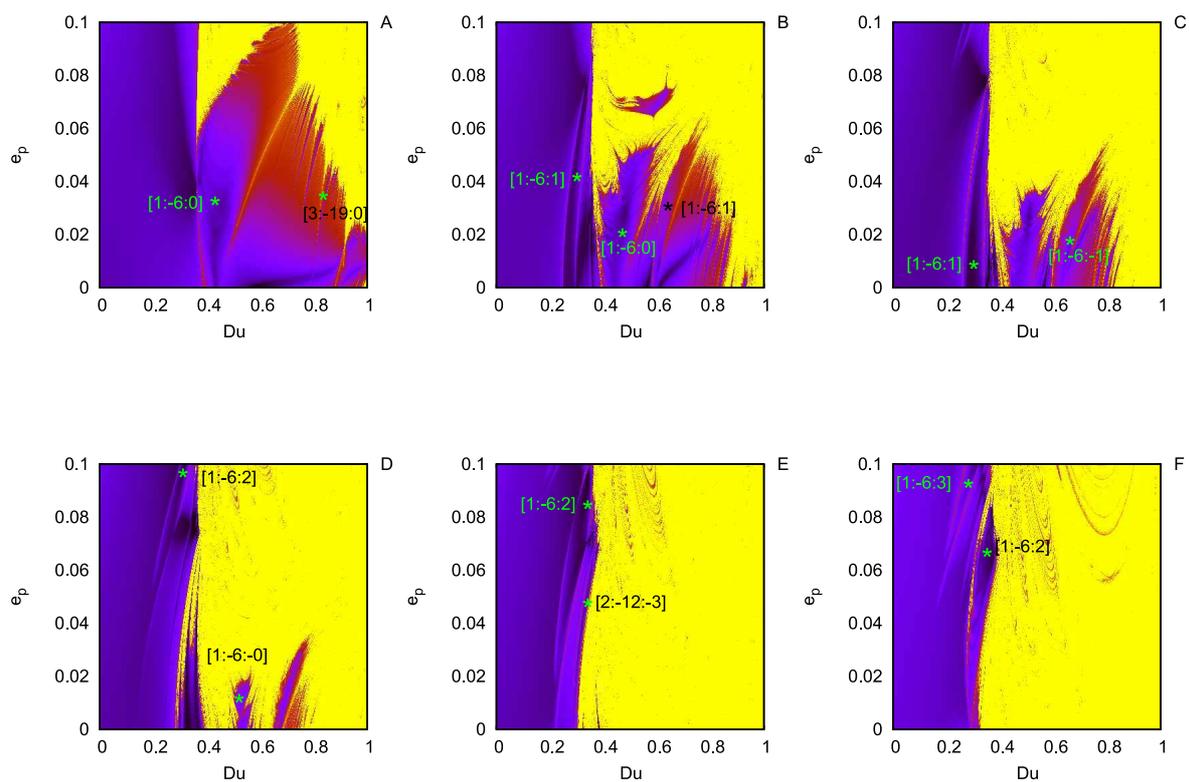}
\vspace{0cm}
\caption{FLI maps for the resonance 1:6, $\mu=0.0041$, 
for the values $e^{\prime}=0$ (A), $e^{\prime}=0.02$ (B), 
$e^{\prime}=0.04$ (C), $e^{\prime}=0.06$ (D), $e^{\prime}=0.08$ 
(E) and $e^{\prime}=0.1$ (F).} 
\label{fli1to6}
\end{figure}

\begin{figure}
\vspace{0cm}
\centering
\includegraphics[width=1.0\textwidth]{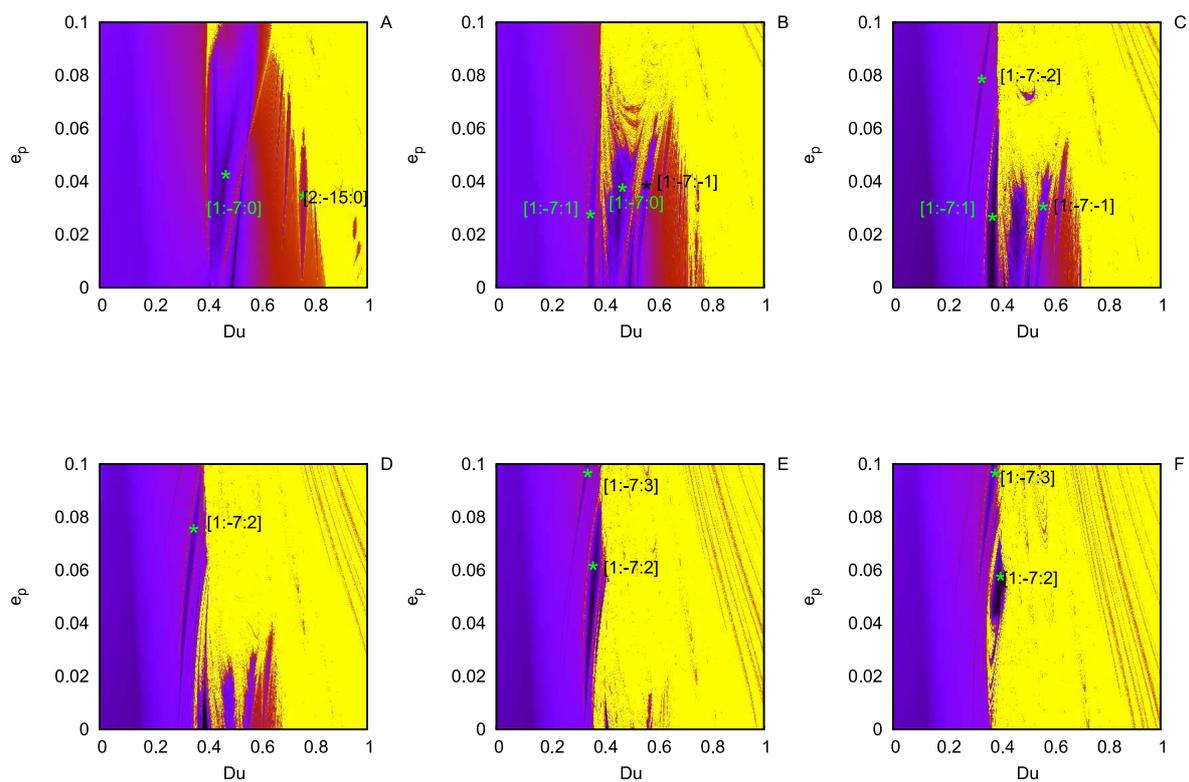}
\vspace{0cm}
\caption{FLI maps for the resonance 1:7, $\mu=0.0031$, 
for the values $e^{\prime}=0$ (A), $e^{\prime}=0.02$ (B), 
$e^{\prime}=0.04$ (C), $e^{\prime}=0.06$ (D), $e^{\prime}=0.08$ 
(E) and $e^{\prime}=0.1$ (F).} 
\label{fli1to7}
\end{figure}

\begin{figure}
\vspace{0cm}
\centering
\includegraphics[width=1.0\textwidth]{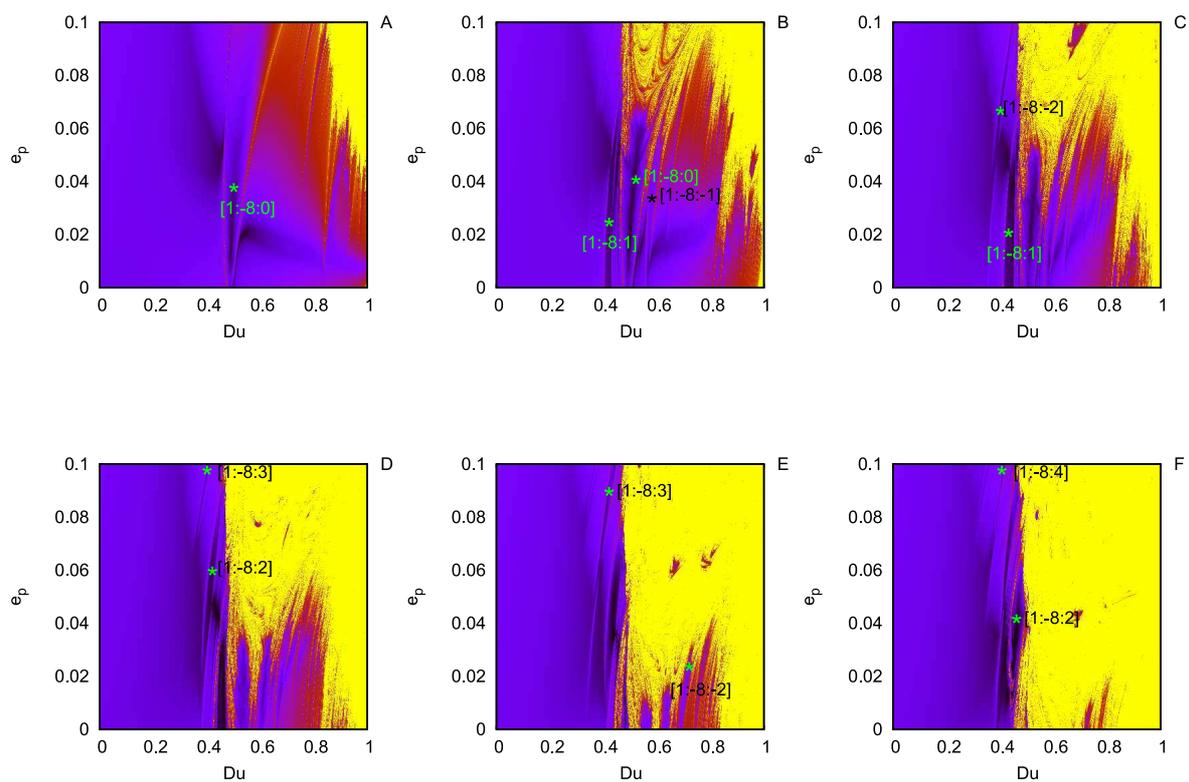}
\vspace{0cm}
\caption{FLI maps for the resonance 1:8, $\mu=0.0024$,
for the values $e^{\prime}=0$ (A), $e^{\prime}=0.02$ (B), 
$e^{\prime}=0.04$ (C), $e^{\prime}=0.06$ (D), $e^{\prime}=0.08$ 
(E) and $e^{\prime}=0.1$ (F).} 
\label{fli1to8}
\end{figure}

\begin{figure}
\vspace{0cm}
\centering
\includegraphics[width=1.0\textwidth]{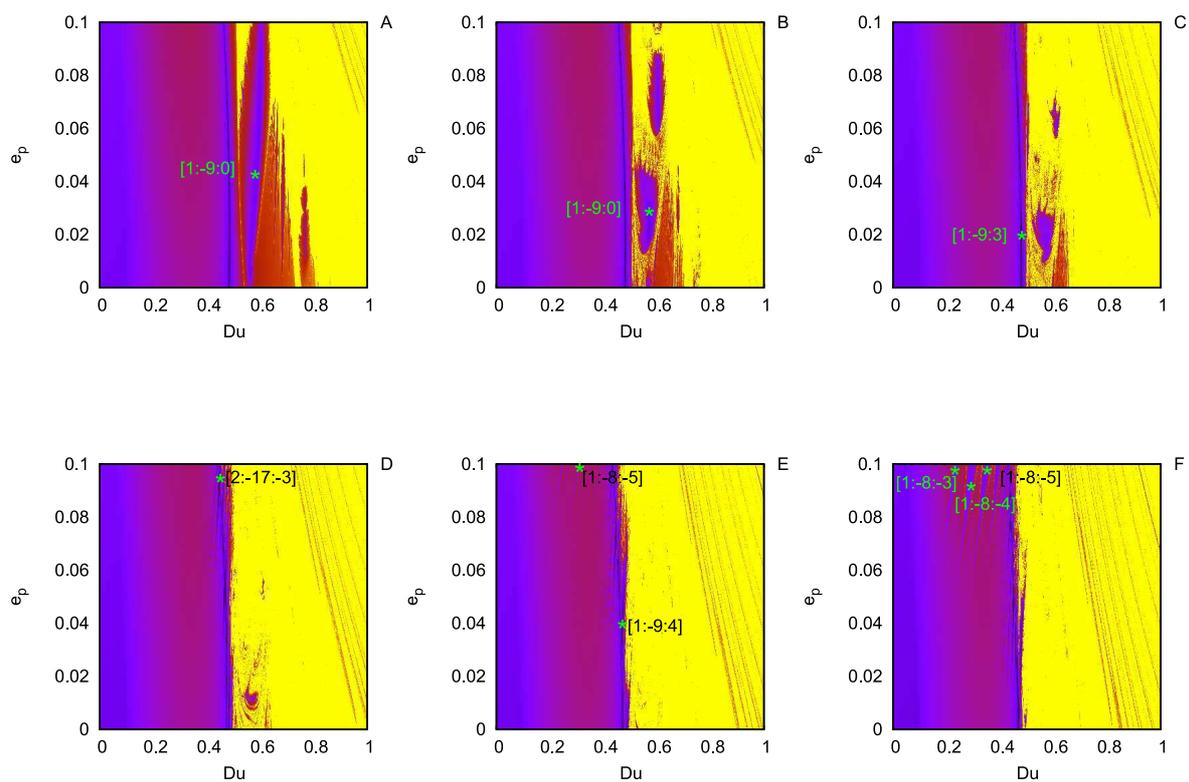}
\vspace{0cm}
\caption{FLI maps for the resonance 1:9, $\mu=0.0021$, 
for the values $e^{\prime}=0$ (A), $e^{\prime}=0.02$ (B), 
$e^{\prime}=0.04$ (C), $e^{\prime}=0.06$ (D), $e^{\prime}=0.08$ 
(E) and $e^{\prime}=0.1$ (F).} 
\label{fli1to9}
\end{figure}

\begin{figure}
\vspace{0cm}
\centering
\includegraphics[width=1.0\textwidth]{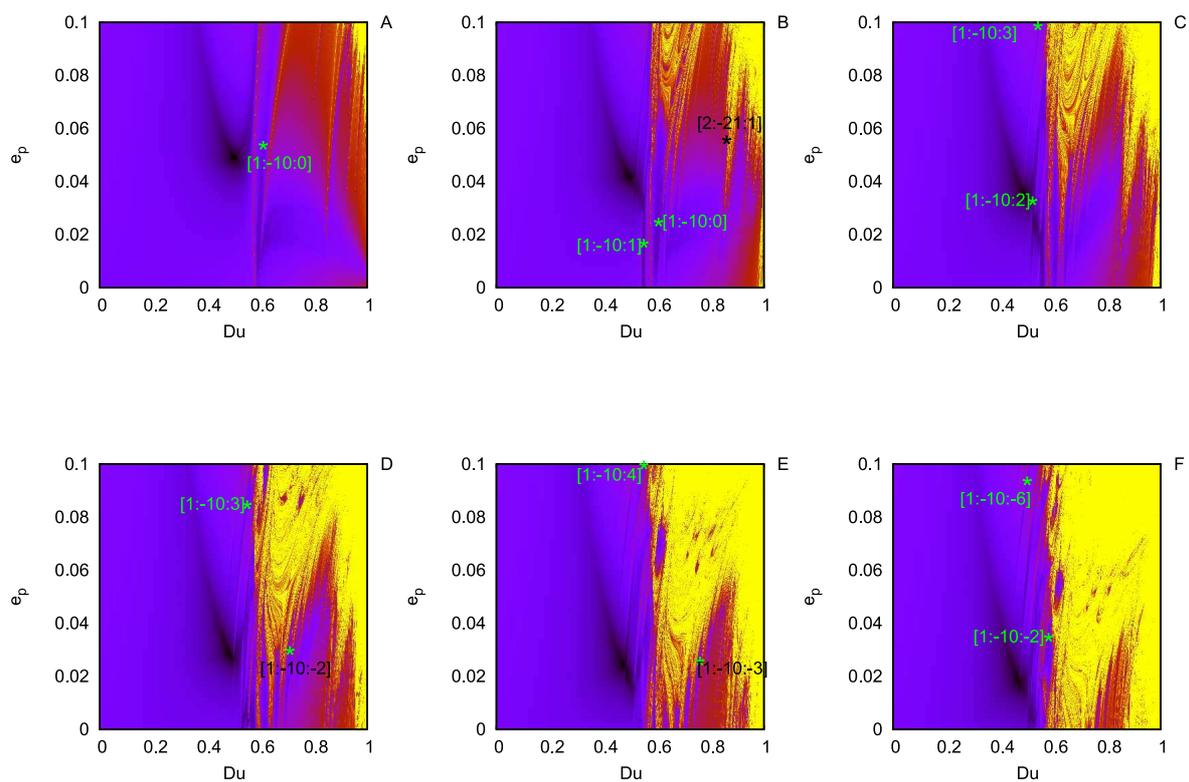}
\vspace{0cm}
\caption{FLI maps for the resonance 1:10, $\mu=0.0016$,
for the values $e^{\prime}=0$ (A), $e^{\prime}=0.02$ (B), 
$e^{\prime}=0.04$ (C), $e^{\prime}=0.06$ (D), $e^{\prime}=0.08$ 
(E) and $e^{\prime}=0.1$ (F).} 
\label{fli1to10}
\end{figure}

\begin{figure}
\vspace{0cm}
\centering
\includegraphics[width=1.0\textwidth]{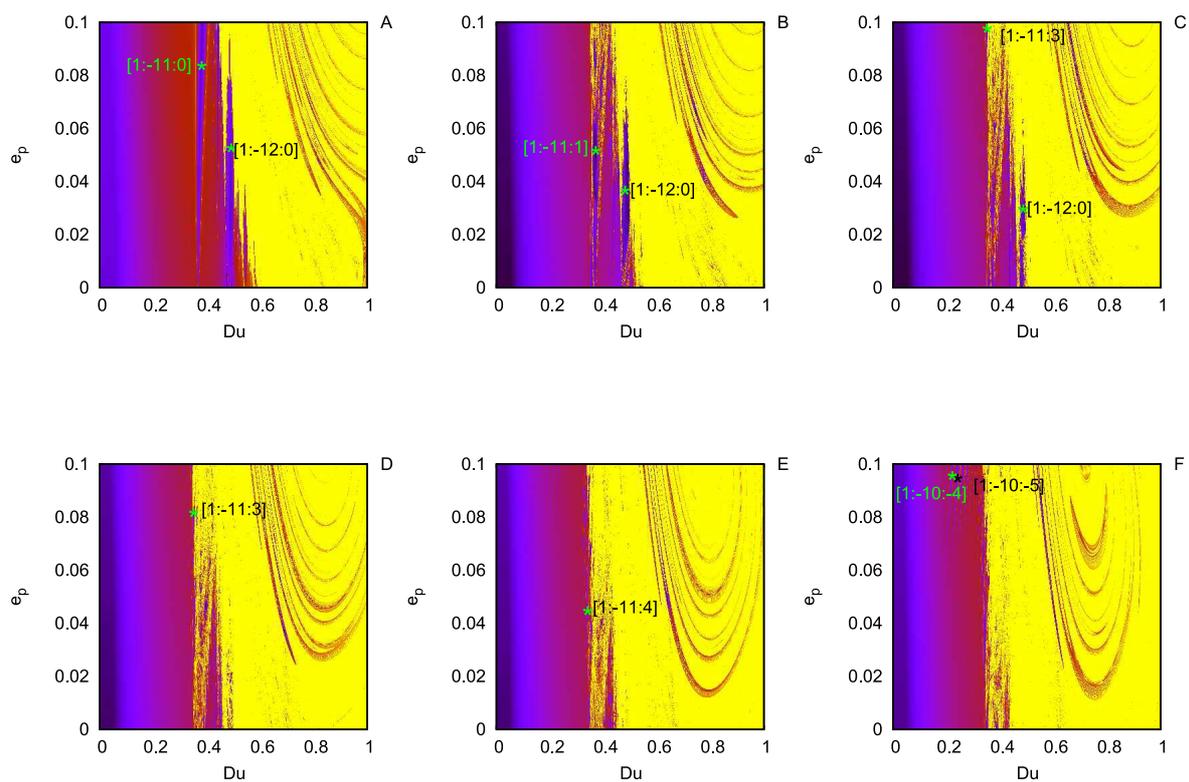}
\vspace{0cm}
\caption{FLI maps for the resonance 1:11, $\mu=0.0014$, 
for the values $e^{\prime}=0$ (A), $e^{\prime}=0.02$ (B), 
$e^{\prime}=0.04$ (C), $e^{\prime}=0.06$ (D), $e^{\prime}=0.08$ 
(E) and $e^{\prime}=0.1$ (F).} 
\label{fli1to11}
\end{figure}

\begin{figure}
\vspace{0cm}
\centering
\includegraphics[width=1.0\textwidth]{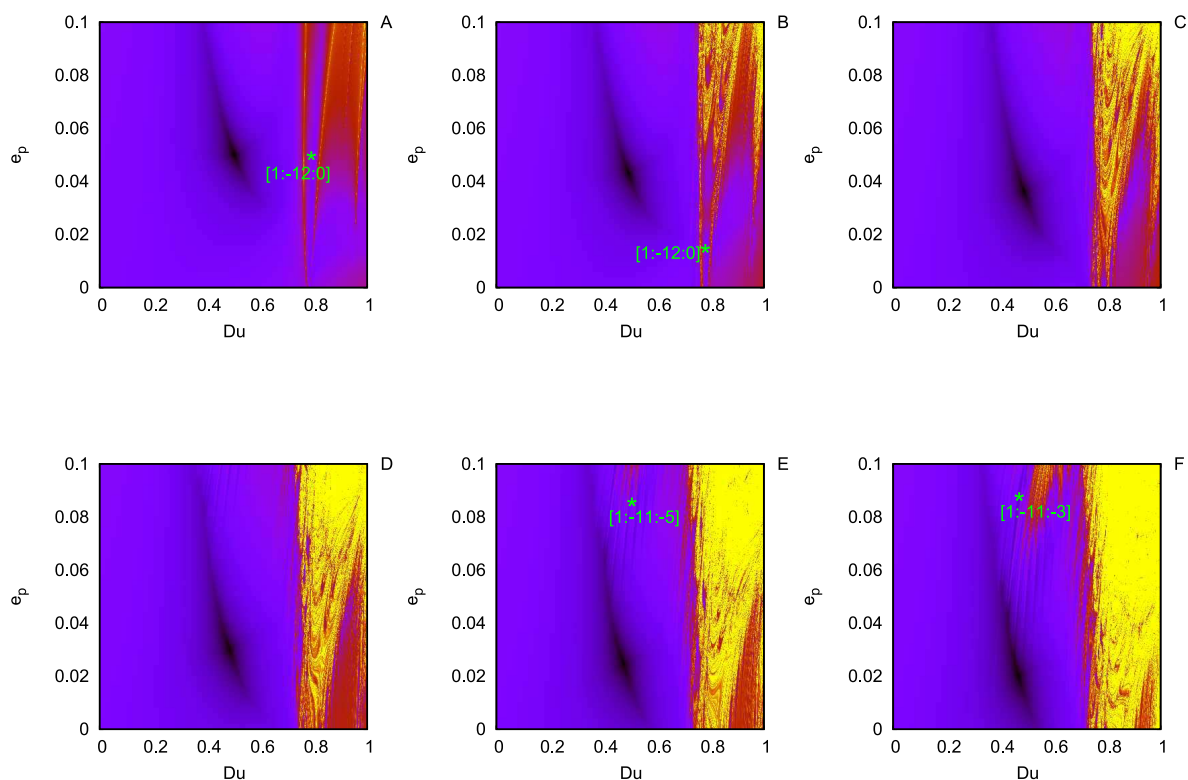}
\vspace{0cm}
\caption{FLI maps for the resonance 1:12, $\mu=0.0012$,
for the values $e^{\prime}=0$ (A), $e^{\prime}=0.02$ (B), 
$e^{\prime}=0.04$ (C), $e^{\prime}=0.06$ (D), $e^{\prime}=0.08$ 
(E) and $e^{\prime}=0.1$ (F).} 
\label{fli1to12}
\end{figure}

\begin{figure}
\hspace{-1.5cm}
\begin{center}
\includegraphics[width=0.95\textwidth,angle=0]{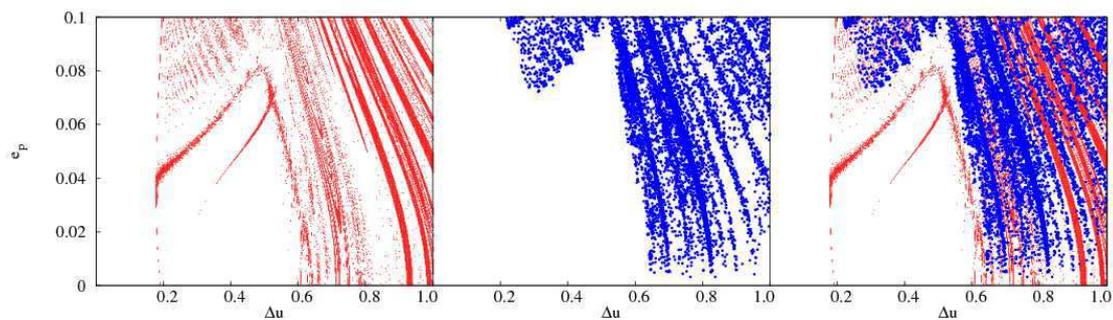}
\end{center}
\caption{Left panel: All initial conditions in the 400$\times$400 
grid of Fig.\ref{fli1to5}A for which the FLI value at the end of 
the integration is in the range $3.5\leq\Psi\leq 6$. Middle panel: 
the points of intersection with the same section as for the FLI 
maps (see text) of the stable invariant manifolds of the family 
of short-period horizontal Lyapunov orbit around L3, computed for 
400 different values of the Jacobi constant as indicated in the 
text. Right panel: superposition of the left and middle panels.}
\label{flivsman}
\end{figure}

Figures \ref{fli1to5} to \ref{fli1to12}, show a surveys of the
resonances, as depicted by numerically computed FLI maps, for
differents cases of $e'$, and for the mass parameters $\mu=0.0012$,
$0.0014$, $0.0016$, $0.0021$, $0.0024$, $0.0031$, $0.0041$ and
$0.0056$. These are representative of the multiplets formed around the
resonances 1:5, 1:6, 1:7, 1:8, 1:9, 1:10, 1:11 and 1:12,
respectively. In each plot, the values for $e^{\prime}$ are $0$,
$0.02$, $0.04$, $0.06$, $0.08$ and $0.1$, from top to bottom and left
to right.  Inspecting these plots, we emphasize the following
features:

i) The size of the non-resonant domain does not change much with
variations of $e^{\prime}$. In fact, in all these plots we observe
that, despite the fact that as $e^{\prime}$ evolves new transversal
resonances appear, the non-resonant domain keeps its limits nearly
constant (at about $0.2$ in Fig \ref{fli1to5}, $0.3$ in Fig
\ref{fli1to6}, $0.4$ in Fig \ref{fli1to7}, $0.45$ in Fig \ref{fli1to8}
and Fig \ref{fli1to9}, $0.6$ in Fig \ref{fli1to10}, $0.35$ in Fig
\ref{fli1to11} and $0.7$ in Fig \ref{fli1to12}). On the other hand,
the resonant domain, which for low (but non-zero) values of
$e^{\prime}$ is filled with small transverse resonances, gradually
shrinks within the chaotic sea, and for values of $e^{\prime}$ around
$0.1$, it completely disappears. This gives a natural limit for the
values of $e^{\prime}$ to consider, since no important resonances
survive for $e'>0.1$.

ii) Around a main resonance, we identify new emerging transverse 
resonances, of the kind $(1,-n,m)$, with $m$ a small integer.  
Since they are triple resonances with commensurability with $g$, 
they are not present for $e^{\prime}=0$, but for greater values 
they become evident, especially some isolated ones which penetrate 
inside the non-resonant region. As $e^{\prime}$ increases, the whole 
structure moves outwards (towards increasing values of $\Delta u$) 
and also downwards (towards smaller values of $e_p$). For bigger 
values of $e^{\prime}$, the main resonances 1:n generally disappear 
or they are small, leaving space for transverse resonances
to dominate in action space.

iii) Finally, in some plots (as in Fig. \ref{fli1to5}, \ref{fli1to6},
\ref{fli1to7}, \ref{fli1to11}), we can see traces, appearing as thin
darker lines in the chaotic domain, of the stable invariant manifolds
emanating from lower-dimensional invariant objects around $L_3$, such
as the short-period planar Lyapunov orbits in the case $e'=0$, or
their associated 2D-tori, for $e'=0$.  Figure \ref{flivsman} shows an
example of comparison of the structures found in the FLI maps with the
exact computation of the stable invariant manifolds of the Lyapunov
orbit around L3 in the case $\mu=0.0056$, $e'=0$, corresponding to
panel A of Fig.\ref{fli1to5}. The left panel shows the same structures
in greater detail, plotting in red all points for which the FLI is in
the limit $3.5\leq \Psi\leq 6$. These limits exclude all points
corresponding to regular orbits, as well as all escaping orbits, for
which the FLI evaluation quickly saturates to a high value $\Psi\geq
50$. On the other hand, the middle panel is computed as follows. We
consider the interval of Jacobi constant values $C_{min}\leq C\leq
C_{max}$, where $C_{min}=2.984$ and $C_{max}=3.00385$ represent the
minimum and maximum value of the Jacobi constant encountered in the
whole 400$\times$400 grid of initial conditions for which the FLI map
of Fig.\ref{fli1to5}A was computed. Splitting this interval in 400
values of $C$, for each value we compute numerically the corresponding
horizontal Lyapunov orbit around L3 and its stable manifold and we
collect all the points in which the latter intersects the section of
the FLI map (given by the pericentric condition
$\lambda'-\omega-\pi/3=0$ as well as $x-0.03\Delta u=0$). Numerically,
we introduce some tolerence $2\times 10^{-4}$ in the section
determination, in order to collect a sufficiently large number of
points necessary for visualization of the results. Plotting in the
same co-ordinates as for the FLI map the collected points for all 400
stable invariant manifolds corresponding to the 400 different values
of C yields the middle plot of Fig.\ref{flivsman}. The relative loss
of sharpness in the picture of the manifolds is due to the numerical
tolerance used in their section's determination.  Despite this effect,
we see clearly that the structures formed by the invariant manifolds
follow in parallel those indicated by the corresponding FLI map. The
possibility to use the FLI (with a small number of iterations), in
order to visualize invariant manifolds was pointed out in Guzzo and
Lega (2013). Here, this effect can be considered as a manifestation of
the so-called `Sprinkler' algorithm (see Kovacs and Erdi 2009 for a
review). Namely, in a system of fast escapes, plotting the initial
conditions of the orbits which have relatively large (forward or
backward) stickiness times (and, thus, relatively lower FLI values
with respect to the escaping orbits) allows to vizualize the (stable
or unstable) manifolds of nearby periodic orbits. In fact, the sticky
chaotic orbits in the forward sense of time are those trapped {\it
  within the lobes} defined by the stable invariant manifolds (see,
for example, figure 19 of Efthymiopoulos et al. 1997). This effect is
clearly shown in our example by combining the left and middle panels
of Fig.\ref{flivsman}. Figure \ref{flivsman} clearly shows that the
points of greater stickiness in the forward sense of time, as revealed
by their relatively low (with respect to fast-escaping orbits) FLI
values, are located precisely between the limits of the structures
indicated by the stable invariant manifolds of the family of the
planar Lyapunov orbits around L3. We note, finally that when $e'>0$,
instead of the foliation of all the manifolds of the Lyapunov family,
one has to consider the invariant manifolds of a 2D unstable invariant
torus around L3. This computation is numerically hardly
tractable. Nevertheless, simple inspection of all panels of
Fig.\ref{fli1to5} clearly shows that the structures found for $e'=0$
essentially continue to exist, in a quite similar geometry, in the
case $e'\neq 0$ as well.


\begin{figure}
\vspace{0cm}
\centering
\includegraphics[width=0.65\textwidth,angle=0]{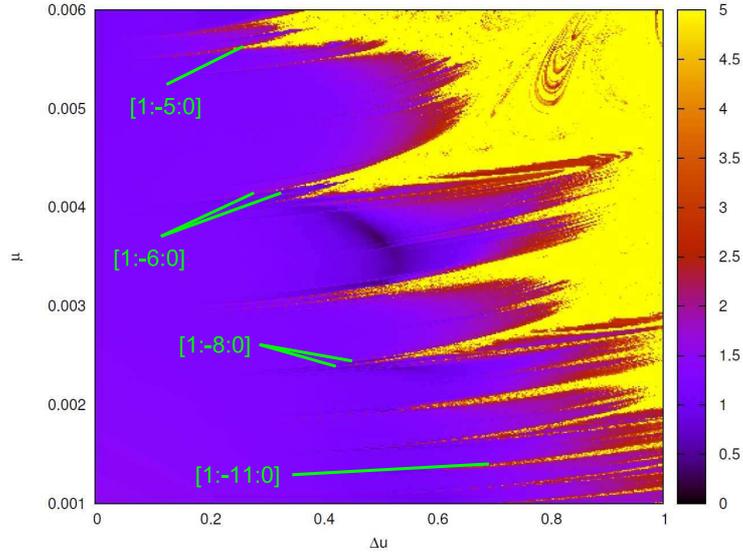}
\caption{FLI map for a grid of $400\times400$ initial
conditions in the plane ($\mu$,$\Delta u$), for $e'=e_p=0.02$ (see text). 
Dark colors (black) correspond to $\Psi=0$ (regular orbits) and light color
 (yellow) to $\Psi=5$ (chaotic orbits). The most important main 
resonances are labeled in the plot. Some compact bundles formed by 
transverse resonances near the main ones are also visible in the 
plot.}
\label{mufli}
\end{figure}

Returning to the discussion of the resonant structure, the overall 
effect of resonances on the size of the stability domain
is resumed in Fig. \ref{mufli}. For a set of initial values of
$\mu$, from $\mu = 0.001$ to $\mu = 0.006$, with step $\Delta \mu =
1.25\times 10^{-5}$, and fixed values for $e^{\prime} = 0.02$ and $e_p
= 0.02$, the figure shows the FLI map produced by integrations of the
set of initial conditions given by $x=0$, $\phi_f=-u-\frac{\pi}{3}$,
$Y_f=0$ and $\Delta u = u_0 + u$ varying from $0$ to $1$, with step
$0.0025$.  Darker (black) colors correspond to regular orbits, and
light (yellow) colors to chaotic orbits. The main resonances appear as
long yellow tongues that contain single or double thin chaotic layers
associated to the separatrix (depending on whether the resonance is
odd or even). In a similar way, a large number of smaller transverse
resonances fill the space between the main ones. By the fact that the
tongues are nearly horizontal, we can infer that the presence of
particular resonances is highly localized with respect to the value of
$\mu$, e.g. the resonance 1:6 is important at $\mu=0.004$, it
completely disappears in the chaotic domain at $\mu=0.0045$.  Also, as
$\mu$ increases, a bifurcation of new secondary resonances
happens less frequently. Nevertheless, since they are of decreasing order,
their width and relative influence increases.

\section{Chaotic diffusion}

The co-existence of different types of resonances renders non-trivial
the question in which domains of the phase space the chaotic
diffusion, due to the interaction of resonances, provides a more
efficient transport mechanism for orbits, thus affecting long term
stability. As explained in section 2, two main regimes of chaotic
transport exist. For isolated resonances located inside the boundary
of the main stability domain (like the transverse resonances
$(1,-6,1)$ and $(1,-6,2)$ of Fig.\ref{freqanal}), the orbits in the
stochastic layer have the possibility of slow diffusion that bears
features of Arnold diffusion. In any case, we find that the diffusion
rate is extremely small, thus it is practically undetectable.  On the
other hand, for resonances located beyond the boundary of the main
stability domain, the diffusion process is best described by the
paradigm of modulational diffusion. In particular, the amplitude of
pulsation of the separatrix-like chaotic layers at the borders of the
resonances is large enough to allow for communication of the
resonances, causing the orbits to undergo abrupt {\it jumps} from one
resonance to another, and eventually to escape.

\begin{figure}
\vspace{0cm}
\centering
\includegraphics[width=1.0\textwidth]{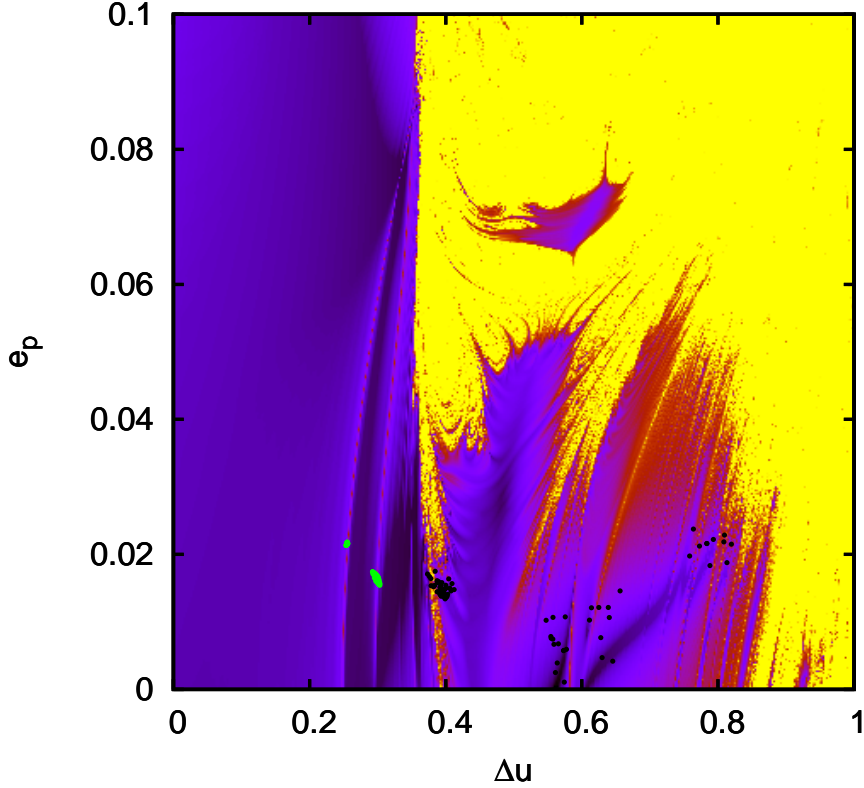}
\caption{Different diffusion processes for two orbits with parameters 
$e^{\prime}\,=\,0.02$, $\mu\,=\,0.0041$, $e_p\,=\,0.01625$ and initial 
conditions $x\,=\,0$, $\phi\,=\,\pi/3$, $Y_f\,=\,0$ and $\Delta u\,=\,0.299$ 
for the green orbit and $\Delta u\,=\,0.376$ for the black orbit.}
\label{flimoddiff}
\end{figure}
Figure \ref{flimoddiff} provides evidence of the processes mentioned
above. Two orbits are shown superposed to the FLI map for
$\mu\,=\,0.0041$, $e'=0.02$. The initial conditions for both
correspond to $e_p=0.01625$, but different $\Delta u$ ($\Delta
u=0.299$, orbit with green points, and $\Delta u=0.376$, orbit with
black points). We plot the intersection points of both orbits with the
plane $(\Delta u,e_p)$ when $x=0$, $\phi_f=0$, with a tolerance
$0.001$ and $0.06$ respectively. According to these data, the first
orbit resides in the chaotic layer of the resonance $(1-6,1)$, while
the second is initially in the chaotic layer of the resonance
1:6. However, the first orbit is restricted to move essentially
only along the stochastic layer of the initial resonance, as no
resonance overlapping exists with any low-order adjacent resonance. As
a result, the orbit's diffusion is practically unobservable. By
contrast, the second orbit suffers a significant change of topology
over a timescale of only $10^5$ periods. The orbit visits many other
resonances besides the starting one, jumping stochastically between
the chaotic layers of the resonances 1:6. $(1,-6,-1)$,
$(1,-6,-2)$, $(3,-19,2)$, and $(3,-19,1)$, and possibly other ones of
higher order. The proper eccentricity $e_p$ also exhibits abrupt jumps
in the interval [$0.001$,$0.025$].

\begin{figure}
\vspace{0cm}
\centering
\includegraphics[width=0.7\textwidth]{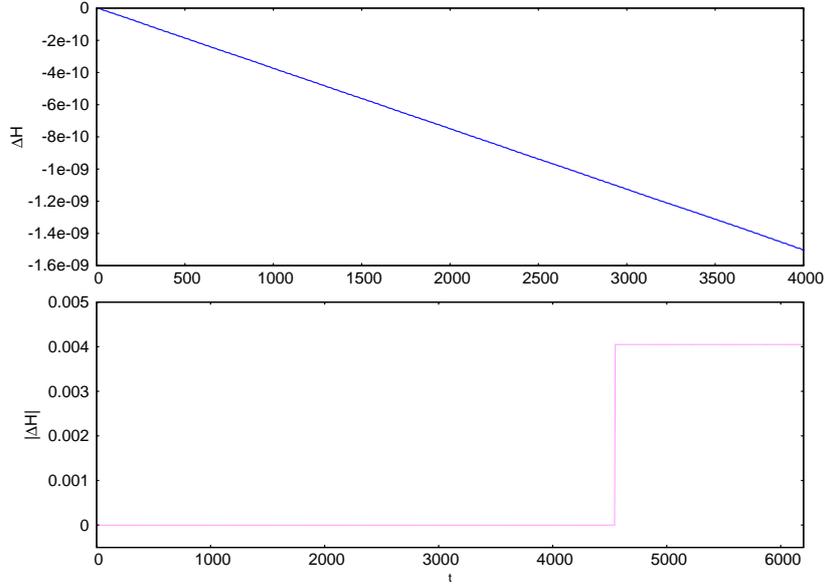}
\caption{Time evolution of the value of the round-off cummulative energy 
error $\Delta H$ of an orbit that escapes at $t\sim 4600$.}
\label{enererr}
\end{figure}

The long-term behavior of orbits in the modulational diffusion regime 
can be characterized by a statistical study. To this end, we consider 
an ensemble of orbits in a rectangle of initial conditions. As an 
example, setting, as before, $\mu=0.0041$, $e'=0.02$, we consider a 
$60\times 60$ grid of initial conditions in the interval $0.33\leq 
\Delta u\leq 0.93$, $0\leq e_p\leq 0.06$, with the remaining initial 
conditions defined as for the FLI maps above. The ensembles were 
processed at 5 different snapshots, corresponding to the integration 
times of $T=10^3$, $10^4$, $10^5$, $10^6$ and $10^7$ periods. In every 
snapshot (of final time $T$), the orbits are classified in three distinct 
groups:\\
\\
\noindent
{\it Regular orbits}: these are orbits whose value of the 
FLI satisfies the condition
\begin{equation}\label{flilim}
\Psi(T)<\log_{10}(\frac{N}{10})
\end{equation}
where $N$ is the total number of periods for the integration. Since for 
regular orbits the FLI grows linearly with $N$, the threshold of Eq.(\ref{flilim}) 
allows to identify orbits which can be clearly characterized as regular. 
These orbits are exempt from further integration.\\
\\
\noindent
{\it Escaping orbits:} an orbit is considered as escaping if the orbit
undergoes a sudden jump in the numerical energy error $\Delta H$
greater than $10^{-3}$. This threshold is determined by the
requirement that the jump surpasses by about two orders of magnitude
the worst possible accumulation of round-off energy errors at the end
of the integration time (i.e. after $10^7$ periods). We tested the
cummulative energy error as a function of time for different initial
conditions. Figure \ref{enererr} shows the evolution of $\Delta H$ for
one example of escaping orbit.  The first panel shows the increment of
$\Delta H$ up to a time $t=4600$.  The absolute cummulative error
grows linearly in time at a rate $\sim 4\times 10^{-13}$~per
period. This rate is characteristic of the orbits in the thin chaotic
layers between the resonances.  However, $\Delta H$ exhibits an abrupt
variation $\Delta H=4\times 10^{-3}$ at the moment of escape. Up to
the maximum integration time $10^7$, the cummulative energy error for
non-escaping orbits is smaller than $4\times 10^{-6}$. Thus, we set a
safe threshold value for escape identifications as $\Delta
H_{esc}=10^{-3}$.\\ \\
\noindent
{\it Transition orbits}: we characterize as transition orbits those whose 
FLI value violates condition (\ref{flilim}), but which do not escape during 
the integration up to the time $T$. As we will see, part of these orbits 
remain at low FLI values up to the and of the integration, yielding 
a growth $\Psi\sim\log(T)$. Thus, the orbits exhibit a regular 
behavior up to at least $10^7$ periods. However, a second sub-group is 
formed among the transition orbits, containing truly sticky orbits 
with positive Lyapunov exponents and FLI values growing asymptotically 
linearly with $T$. \\
\\
With the results at the five different snapshots, a statistical 
study of the escaping times is constructed as follows: in the end of every 
snapshot, 
i) we count the number of orbits belonging to each of the three groups, 
ii) we compute the histogram of FLI values (from $0$ up to $50$) for the 
transition orbits, and iii) we store the values of $\Delta H$ and $\Psi$ 
for both the escaping and the transition orbits. The results of (i) are 
summarized in the following table:\\
\begin{center}
\begin{tabular}{ccccc}
\hline
Snapshot &(N. of periods) &Regular &Transition &Escaping\\
\hline
\hline
1 &$10^3$ &1220 ($33.8$\%) &2027 ($56.3$\%) &353 ($9.9$\%) \\
\hline
2 &$10^4$ &1263 ($35$\%) &1388 ($38.5$\%) &949 ($26.5$\%) \\
\hline
3 &$10^5$ &1296 ($36$\%) &966 ($26.8$\%) &1338 ($37.2$\%) \\
\hline
4 &$10^6$ &1299 ($36.1$\%) &699 ($19.4$\%) &1602 ($44.5$\%) \\
\hline
5 &$10^7$ &1309 ($36.3$\%) &603 ($16.8$\%) &1688 ($46.9$\%) \\
\hline
~&~&~&~&~\\
\end{tabular}
\end{center}

\begin{figure}
\centering
\includegraphics[width=0.9\textwidth]{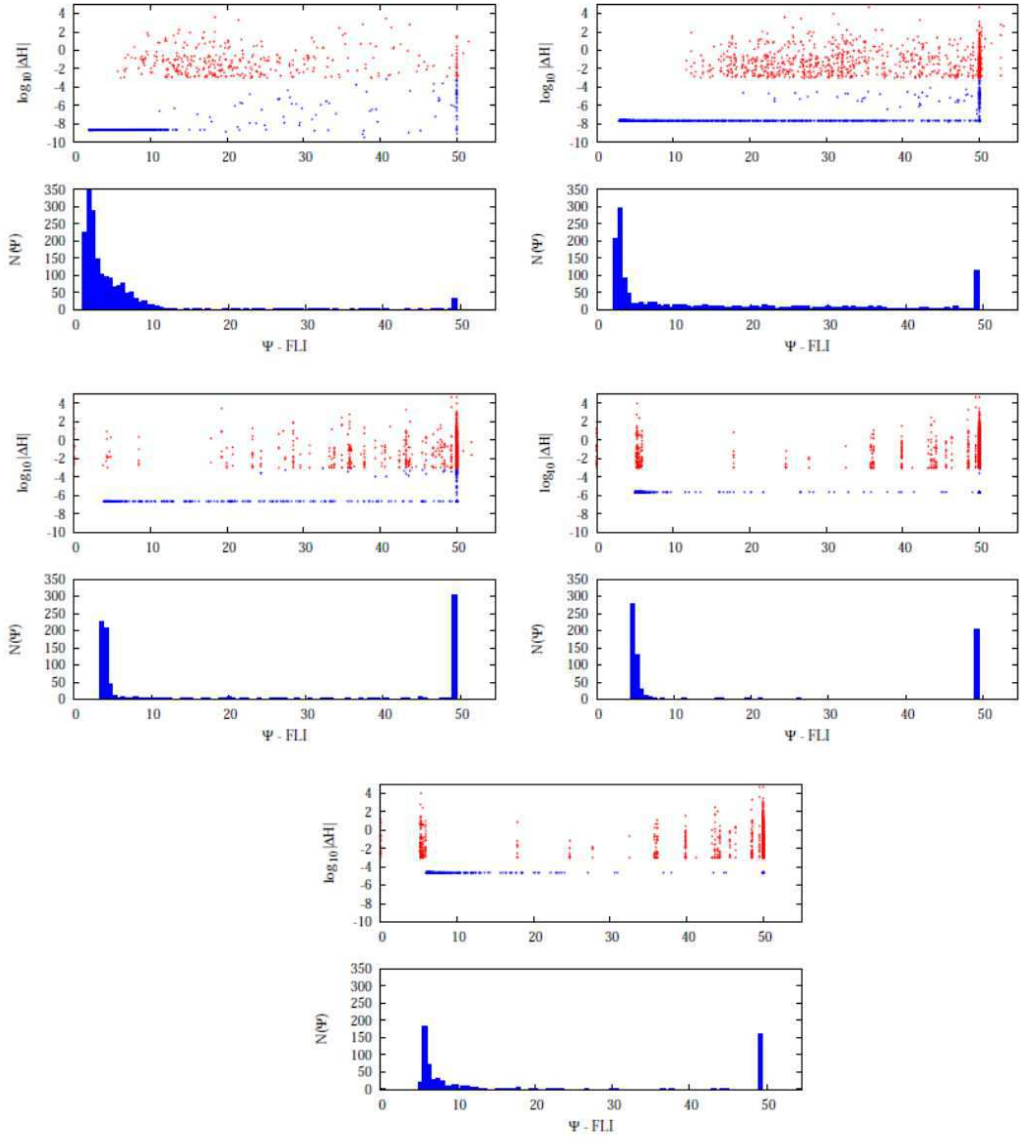}
\caption{Cummulative energy round-off error $\Delta H$ vs FLI value  for 
the groups of transitions (blue) and escaping (red) orbits, as well as the 
distribution of the FLI values for the transition group. The different
panels refer to the time snapshots $T\,=\,10^3$ (a), 
$T\,=\,10^4$ (b), $T\,=\,10^5$ (c), $T\,=\,10^6$ (d) and $T\,=\,10^7$ (e) 
periods of the primaries.}
\label{histo}
\end{figure}
Focusing, now, on the groups of transition and escaping orbits, the
upper panels in Fig. \ref{histo}a-e show the distribution of $\Delta
H$ and FLI ($\Psi$) values for the transition orbits (blue) and the
escaping orbits (red) respectively. In Fig.\ref{histo}a (upper panel),
for $T=10^3$ periods, most of the transition orbits are found to keep
a relatively low value of the FLI, $\Psi<10$, and a cummulative energy
error $\Delta H$ of about $10^{-9}$. In fact, {\it all} the transition
orbits with larger $\Delta H$ become escaping orbits shortly after
$T=10^3$ periods. A second group of transition orbits, however, starts
being formed, with FLI larger than or equal to 50. The lower panel
shows the distribution of FLI values for all the transition
orbits. The left concentration represents regular or very sticky
orbits, while the more chaotic orbits are spread over larger values of
the FLI, with a small secondary peak formed in the right part of the
histogram at $\Psi=50$.  However, as the integration time increases, a
`stream' is formed that transports members of the left group towards
the right group. As a result, at the last snapshot, ($T=10^7$
periods), the right group contains about 30\% of the transition orbits
and 6\% of the total orbits considered. In fact, as visually clear in
all upper panels of Fig.\ref{histo}, most escapes occur at
intermediate values of the FLI, while the right group is nearly
completely detached from the left group of the transition orbits, the
latter moving to the right at a speed logarithmic in $T$, i.e. as
expected for regular orbits. Finally, the Lyapunov characteristic
times of the orbits in the right group are all substantially smaller
than $T_L=10^5$, while the orbits remain sticky for times
$T_{stickiness}> 2\pi 10^7$.  This behaviour is reminiscent of
\emph{stable chaos}.

Finally, we notice that the escaping orbits (red) seem to form bands
of preferential values of the FLI. We have not fully identified the
origin of these bands. Nevertheless, they could be connected to the
fact that the escape can occur only via the thin chaotic layers
between the resonances, so that the concentration to particular FLI
values could reflect the local FLI value for orbits residing for long
time within each one of such layers.

\begin{figure}
\centering
\includegraphics[width=0.5\textwidth,angle=270]{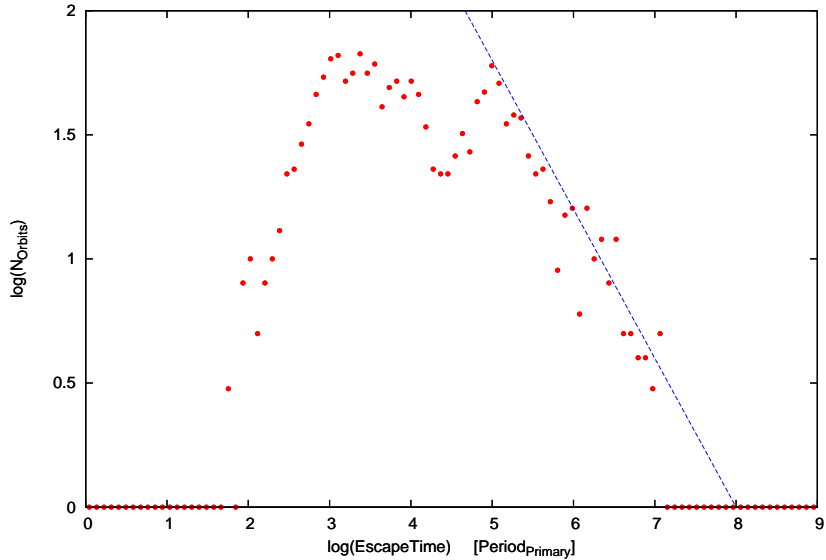}
\caption{Histogram of escaping times for the escaping orbits.}
\label{histoesc}
\end{figure}
Figure \ref{histoesc} shows the histogram of escaping times of all the
escaping orbits. It becomes evident that in the process of escaping
two distinct timescales can be distinguished, corresponding to two
peaks of the histogram. The first peak, at about $10^3$ periods,
corresponds to fast escapes, while the second, at about $10^5$
periods, corresponds to slow escapes. As shown below, the large
majority of fast escaping orbits are with initial conditions within
the chaotic sea surrounding the resonances, while slowly escaping
orbits are those with initial conditions in the thin chaotic layers
delimiting the resonances. In the latter case, we find that beyond a
time $t\approx 10^5$ periods, the distribution of the escape times 
shows a {\it power-law} tail. The straight line in Fig.\ref{histoesc}
represents a power-law fit
\begin{equation}
P(t_{esc})\propto t_{esc}^{-\alpha},~~~~\alpha\approx 0.8~~~.
\end{equation}
We note in this respect that power-law statistics of the escape times
are a characteristic feature of stickiness and long-term chaotic
correlations of chaotic orbits (see Meiss 1992, p. 843 and references 
therein, and also Ding et al 1990, Cheng et al 1992).

\begin{figure}
\centering
\includegraphics[width=0.75\textwidth,angle=270]{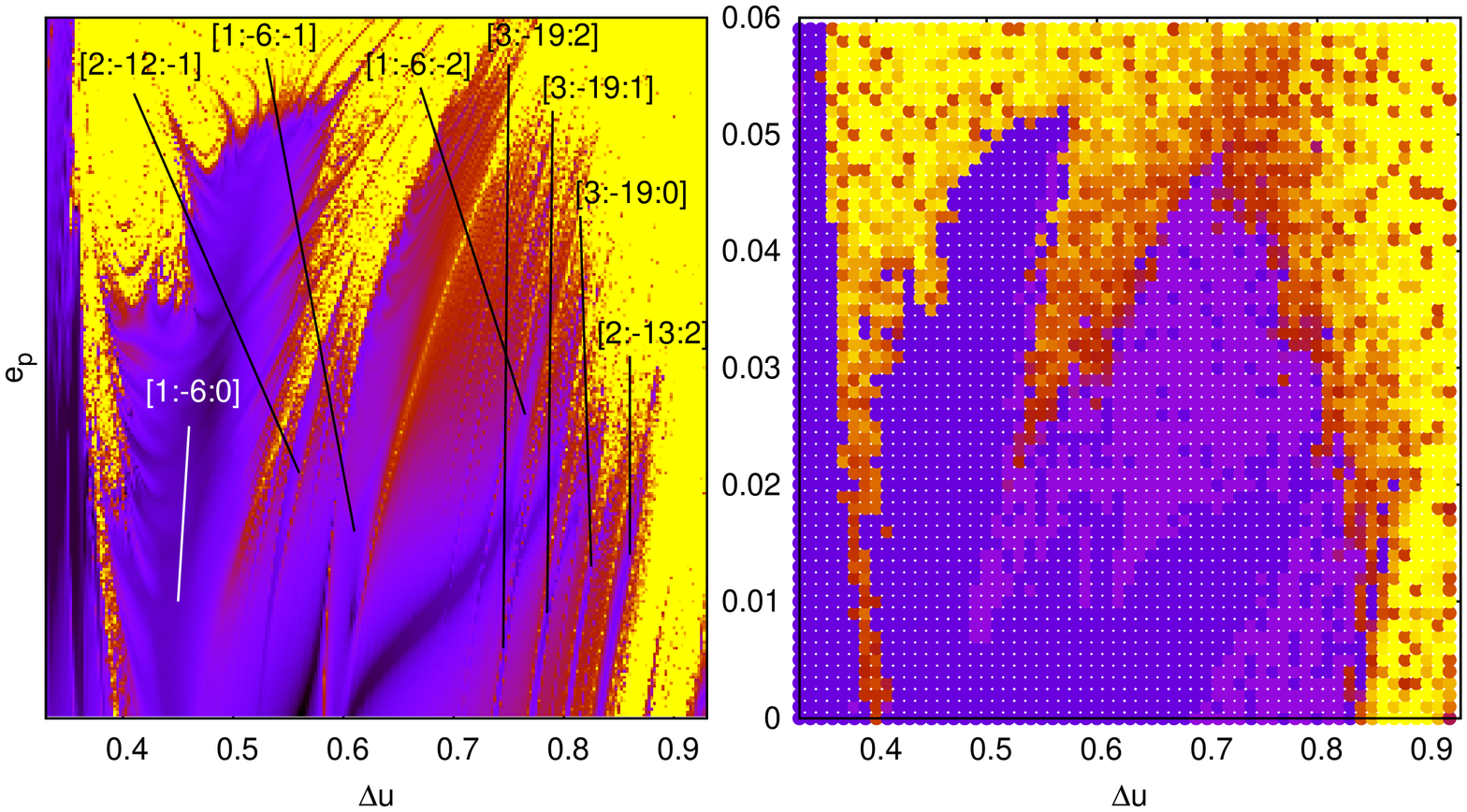}
\begin{center}
\includegraphics[width=0.8\textwidth]{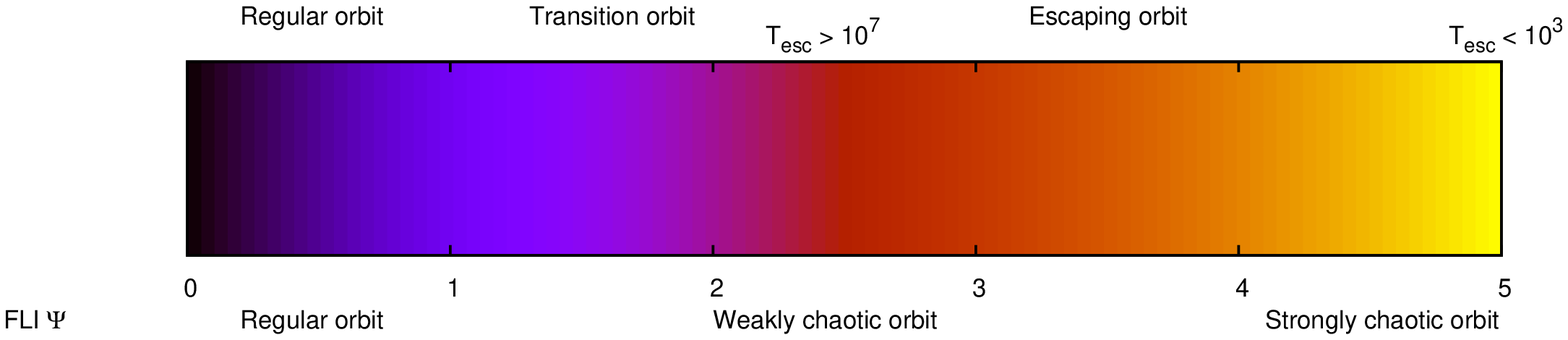}
\end{center}
\caption{FLI map for the grid of initial conditions $[0.33,0.93]\,
  \mathsf{x}\,[0.00,0.06]$ in the plane $(\Delta u,e_p)$ (left panel)
  and color map of the escaping times for the same grid (right panel),
  in both cases with parameters $\mu=0.0041$ and $e'=0.02$. See text
  for more details.}
\label{fliesc}
\end{figure}
Finally, Fig.\ref{fliesc} shows a comparison between FLI values and
the residence (or escape) times for all the orbits of the integrated
ensemble.  The left panel shows the FLI map for the square of initial
conditions $[0.33,0.93]\times [0.00,0.06]$. The right panel, now,
shows in color scale the residence, or escaping times, for all $3600$
initial conditions of the ensemble. Yellow colors represent the faster
escaping times ($t_{esc}<10^3$ periods), while red the slower (between
$10^6$ and $10^7$ periods). In light purple are depicted the orbits
that remain in the group of transition orbits up to the end of the
integration ($10^7$ periods), while deep purple represents the orbits
belonging to the regular group. The first observation is that the
distribution of escaping times reproduces to some extent the main
features of the resonant structure found by the stability map. In
particular, the chaotic layer of the resonance 1:6 appears
clearly marked by long escape times (larger than $10^5$ periods).
The stickiness is in general enhanced at the borders of all
resonances. On the other hand, most of the orbits with $\Psi=5$
or greater, that are qualified as chaotic in the stability map, belong
to the population of fast escapes, and we find that most need less
than $10^3$ periods to escape. But the most interesting feature is
that all the thin chaotic layers around the resonances $(2,-12,-1)$,
$(1,-6,-1)$, $(1,-6,-2)$, $(3,-19,2)$, $(3,-19,1)$, $(3,-19,-0)$, and
$(2,-13,2)$, contain orbits that appear \emph{not escaping} at least up to
$10^7$ periods. This, despite the fact that some of these orbits are
relatively strongly chaotic, i.e with $\Psi$ close to 5. Thus, we can
conclude that the stickiness phenomena in the thin chaotic layers
formed in the resonant domains of the action space can prolong the
stability of hypothetical trojan planets up to times comparable to the
age of the hosting system. Consequently, the resonant domains beyond
the main stability domain are also candidates to host trojan
companions of giant extrasolar planets.

\section{Conclusions}

In this work, we present a parametric study of the resonant
structure in the co-orbital domain, as well as the main types and 
associated timescales for chaotic diffusion of hypothetical
small (considered massless) trojan companions of giant planets 
in extrasolar planetary systems. 

First, we introduce a Hamiltonian formalism in terms of modified
Delaunay action-angle variables, allowing to take various secular
effects into account. In particular, while we presently deal in detail
with the planar ERTBP, the proposed formalism allows for
generalizations in the case of more than one planet, as well as
non-zero inclinations of the exoplantes. This formulation permits us
to define a hierarchy of Hamiltonian models corresponding to different
levels of perturbation on the trojan planetary motion. In particular,
the hamiltonian $H_b$ (eq. (\ref{hambasic2})) represents a system of
two degrees of freedom that describes the synodic (guilding-center)
and short period motions of the trojan planet. Its associated
parameters $J_s$, $Y_p$ serve as constants of motion (proper
elements).  Furthermore, modulo a redefinition of variables, the form
of $H_b$ is identical in the ERTBP and in the RMPP. The full
hamiltonian consists of $H_b \, +$ terms describing the secular
effects due to the primary or to additional perturbing planets.
We classified resonances between the short-period, synodic and secular
frequencies, in three types:
\emph{main}, \emph{transverse} and \emph{secular}.

On the other hand, by means of numerical experiments in the planar
ERTBP, we depict the form of the corresponding web of resonances in
the space of proper elements. With a parametric study for various
values of $\mu$ (in the range 0.001$\leq \mu \leq$0.006) and
$e^{\prime}$ ($e^{\prime} \leq$ 0.1) for the giant
primary, we produce phase portraits and stability maps.

In the phase portraits, we found that the resonant orbits move
outwards as the proper eccentricity $e_p=\sqrt{-2Y_p}$ increases, and
the corresponding stability islands grow. For small values of $e_p$
the stability islands are surrounded by invariant tori.  Small higher
order resonances are present at the border of the stability
domain. Since as $e_p$ increases the resonant islands grow in size,
most of their surrounding invariant tori are destroyed. For a critical
value of $e_p$, the last KAM torus surrounding the resonant island
chain is destroyed, and the chaotic sea penetrate deeper inside the
main stability domain.

Similar phenomena occur by varying the mass parameter $\mu$. In this
case, different values of $\mu$ give rise to different
resonances. Thus, the size of the domain of stability suffers abrupt
variations connected to the bifurcations of new resonances.  On the
other hand, the size of the non-resonant domain does not change much
with variations on $e^{\prime}$. However, the resonant domain, which
for low values of $e^{\prime}$ is full of transverse resonances,
quickly shrinks within the chaotic region, and for values of
$e^{\prime}$ beyond 0.1, it completely disappears.

We also present some main types of chaotic diffusion observed
by the integrations of orbits with initial conditions in the resonance
web. We found two different kinds of diffusion processes, depending 
on the width
of the pulsation of the separatrices. For resonances with small
pulsation widths, or isolated (as the transverse resonances that
penetrate the domain of stability), the separatrices of one resonance
do not enter to the pulsation domain of nearby resonances. Then, the rate
of the chaotic diffusion is practically undetectable. On the other
hand, for resonances with large pulsation widths, or closely spaced 
multiplets, the separatrices of more than one nearby resonances overlap, 
and the proccess of diffusion accelerates, with timescale of
the order of 1Myr. This type of diffusion is best described by
models of modulational diffusion.

Integrating ensembles of orbits, we produce a statistical study of the 
escape probabilities as a function of the escape time. Two distinct 
escaping timescales are identified (fast and slow), characterized by
two peaks in the corresponding histrogram. The slow escapes exhibit
a power-law distribution of the escaping times, which is characteristic 
of stickiness and long-term correlations of weakly chaotic orbits.

Finally, we compare the results of the escaping (or residence) times,
with the results produced by the FLI maps for the same parameter and
set of initial conditions. We conclude that: i) the
distribution of escaping times reproduces to some extent the resonant
structure found by the FLI map. ii) Sticky orbits are located at the
borders of the resonant domain.  Besides, the orbits qualified as
chaotic in the stability map, show fast escapes and need less than
$10^3$ periods for escaping. Finally, iii) the thin chaotic layers in 
the resonant domain contain orbits that appear not escaping at least 
up to $10^7$ periods, despite the fact that most of these orbits are
quite strongly chaotic. As a result, the stickiness phenomena of these 
thin chaotic layers can prolong the stability of hypothetical trojan
planets up to the lifetime of their hosting system.

\vspace{0.5cm}
\noindent
{\bf Acknowledgements:} The authors would like to thank Professors
Z. Knezevic and U. Locatelli for their useful remarks, and the
anonymous referees for helping to improve the original
manuscript. R.I.P. is supported by the Astronet-II Marie Curie
Training Network (PITN-GA-2011-289240).  C.E. acknowledges support by
the Grant 200/815 of the Research Committee of the Academy of Athens.

\newpage
\section{Appendix}

\subsection{Averaged Hamiltonian $<H>$}

Up to the second order in $e$ and $e'_0$, and first order in $\mu$, the 
averaged Hamiltonian $<H>$ of Eq.(\ref{hamsec3}) is given by

$$ <H> = -\frac{1}{2 \left( 1+x \right)^2} - x + J_3 - g' \, y + \frac{\mu}{a} + a\,  e\, e'_0 \, \mu\, \cos(\beta - 2\tau) + a\, \mu\, \cos(\tau) $$

$$ - \frac{1}{2} \, a \, e^2 \, \mu \, \cos(\tau) - \frac{1}{2} \, a \, e^{'2}_0 \, \mu \, \cos(\tau) - \frac{15 \, a^2 \, e^2 \, \mu}{8 \left( 1 + a^2 - 2 \, a \, \cos(\tau) \right)^{5/2}} - \frac{3 \, a^4 \, e^2 \, \mu}{4 \left( 1 + a^2 - 2 \, a \, \cos(\tau) \right)^{5/2}}$$

$$ - \frac{3 \, e^{'2}_0 \, \mu}{4 \left( 1 + a^2 - 2 \, a \, \cos(\tau) \right)^{5/2}} - \frac{15 \, a^2 \, e^{'2}_0 \, \mu}{8 \left( 1 + a^2 - 2 \, a \, \cos(\tau) \right)^{5/2}} + \frac{9 \, a \, e\, e'_0\, \mu \, \cos(\beta)}{4 \left( 1 + a^2 - 2 \, a \, \cos(\tau) \right)^{5/2}} $$

$$ + \frac{9 \, a^3 \, e \, e'_0 \, \mu \, \cos(\beta)}{4 \left( 1 + a^2 - 2 \, a \, \cos(\tau) \right)^{5/2}} - \frac{3 \, a^2\, e \, e'_0 \, \mu \, \cos(\beta - 3 \tau)}{8 \left( 1 + a^2 - 2 \, a \, \cos(\tau) \right)^{5/2}} - \frac{3 \, a \, e \, e'_0 \, \mu \, \cos(\beta - 2 \tau)}{4 \left( 1 + a^2 - 2 \, a \, \cos(\tau) \right)^{5/2}}$$

$$ - \frac{3 \, a^3 \, e \, e'_0 \, \mu \, \cos(\beta - 2 \tau)}{4 \left( 1 + a^2 - 2 \, a \, \cos(\tau) \right)^{5/2}} + \frac{3 \, a^2 \, e \, e'_0 \, \mu \, \cos(\beta - \tau)}{4 \left( 1 + a^2 - 2 \, a \, \cos(\tau) \right)^{5/2}} + \frac{3 \, a^3 \, e^2 \, \mu \, \cos(\tau)}{2 \left( 1 + a^2 - 2 \, a \, \cos(\tau) \right)^{5/2}} $$

$$ + \frac{3 \, a \, e^{'2}_0 \, \mu \, \cos(\tau)}{2 \left( 1 + a^2 - 2 \, a \, \cos(\tau) \right)^{5/2}} + \frac{3 \, a^2 \, e^2 \, \mu}{4 \left( 1 + a^2 - 2 \, a \, \cos(\tau) \right)^{3/2}} + \frac{3 e^{'2}_0 \, \mu}{4 \left( 1 + a^2 - 2 \, a \, \cos(\tau) \right)^{3/2}} $$

$$ - \frac{9 \, a \, e \, e'_0 \, \mu \, \cos(\beta)}{4 \left( 1 + a^2 - 2 \, a \, \cos(\tau) \right)^{3/2}} - \frac{ a \, e \, e'_0 \, \mu \, \cos(\beta - 2\tau)}{4 \left( 1 + a^2 - 2 \, a \, \cos(\tau) \right)^{3/2}} + \frac{a \, e^2 \, \mu \, \cos(\tau)}{2 \left( 1 + a^2 - 2 \, a \, \cos(\tau) \right)^{3/2}} $$

$$ + \frac{a \, e^{'2}_0 \, \mu \, \cos(\tau)}{2 \left( 1 + a^2 - 2 \, a \, \cos(\tau) \right)^{3/2}} - \frac{\mu}{\left( 1 + a^2 - 2 \, a \, \cos(\tau) \right)^{1/2}} + \frac{9 \, a^2 \, e^2 \, \mu \, \cos(2 \tau)}{8 \left( 1 + a^2 - 2 \, a \, \cos(\tau) \right)^{5/2}}$$

$$ \frac{9 \, a^2 \, e^{'2}_0 \, \mu \, \cos(2 \tau)}{8 \left( 1 + a^2 - 2 \, a \, \cos(\tau) \right)^{5/2}} - \frac{27 \, a^2 \, e \, e'_0 \, \mu \, \cos(\beta + \tau)}{8 \left( 1 + a^2 - 2 \, a \, \cos(\tau) \right)^{5/2}} \, ,$$
\\
where $a = (x+1)^2$ and $e = \sqrt{1 - \left( \frac{y}{x+1} + 1\right)^2}$.

\newpage
\subsection{Form of the function $H_b$}

Neglecting ${\cal O}(x)$ terms, and setting $b_0=2-2\cos\tau$,
$e_{p,0}=\sqrt{2(Y_f-Y_p)}$, the functions ${\cal F}^{(0)}$ and ${\cal F}^{(1)}$ of
Eq.(\ref{hamell}), up to second order in $e_{p,0}$ and $e'$, are 
analyzed in trigonometric terms in the angles
$\tau$, $\phi_f$, and $\phi$, as follows:

\vspace{1cm}
\begin{adjustwidth}{-4em}{+3em}
\begingroup
\fontsize{10pt}{12pt}\selectfont
$<{\cal F}^{(0)}>$\\
\begin{center}
\begin{tabular}{||l c || l c ||}
\hline
Constant & $-1+{1\over b_0^{1/2}}+{1\over b_0^{3/2}}\left(-{3e'^2\over 8}-{3e_{p,0}^2\over 4}\right)+{1\over b_0^{5/2}}\left(3e'^2+{21e_{p,0}^2\over 8}\right)$ & $\qquad$ & $\qquad$\\
\hline
$\cos\tau$ & $-1+{e_{p,0}^2\over 2}+e'^2+{1\over b_0^{3/2}}\left(-e'^2-{e_{p,0}^2\over 2}\right)+{1\over b_0^{5/2}}\left(-{27e'^2\over 16}-{3e_{p,0}^2\over 2}\right)$ & $\sin \tau$ & ${1\over b_0^{5/2}}\left(-{33\sqrt{3}e'^2\over 16}\right)$\\
\hline
$\cos 2\tau$ & $-{e'^2\over 2}+{1\over b_0^{3/2}}\left({e'^2\over 8}\right)
+{1\over b_0^{5/2}}\left(-{3e'^2\over 2}-{9e_{p,0}^2\over 8}\right)$ & $\sin 2\tau$ & $\quad -{\sqrt{3}e'^2\over 2}+{1\over b_0^{3/2}}\left({\sqrt{3}e'^2\over 8}\right)+{1\over b_0^{5/2}}\left({3\sqrt{3}e'^2\over 4}\right)$\\
\hline
$\cos 3\tau$ & $+{1\over b_0^{5/2}}\left({3e'^2\over 16}\right)$ & $\sin 3\tau$ & $+{1\over b_0^{5/2}}\left({3\sqrt{3}e'^2\over 16}\right)$\\
\hline
\end{tabular}\\
\end{center}

\vspace{2cm}

\noindent
$\tilde{{\cal F}}^{(0)} = {\cal F}^{(0)}-<{\cal F}^{(0)}>$\\
\begin{center}
\begin{tabular}{||l c || l c ||}
\hline
$\cos(\phi_f)$ & ${3e_{p,0}\over 2}+{1\over b_0^{3/2}}\left(-{3e_{p,0}\over 2}\right)$ &
$\cos(\phi_f+\tau)$ & $-e_{p,0}+{1\over b_0^{3/2}}\left(e_{p,0}\right)$ \\
\hline
$\cos(\phi_f+2\tau)$ & $-{e_{p,0}\over 2}+{1\over b_0^{3/2}}\left({e_{p,0}\over 2}\right)$& $\cos(2\phi_f)$ & ${1\over b_0^{5/2}}\left({27e_{p,0}^2\over 16}\right)$ \\
\hline
$\cos(2\phi_f+\tau)$ & $-{e_{p,0}^2\over 8}+{1\over b_0^{3/2}}\left({e_{p,0}^2\over 8}\right)+{1\over b_0^{5/2}}\left(-{9e_{p,0}^2\over 4}\right)$ & $\cos(2\phi_f+2\tau)$ & $-e_{p,0}^2+{1\over b_0^{3/2}}\left({e_{p,0}^2\over 4}\right)+{1\over b_0^{5/2}}\left(-{3e_{p,0}^2\over 8}\right)$ \\
\hline
$\cos(2\phi_f+3\tau)$ & $-{3e_{p,0}^2\over 8}+{1\over b_0^{3/2}}\left({3e_{p,0}^2\over 8}\right)+{1\over b_0^{5/2}}\left({3e_{p,0}^2\over 4}\right)$ & $\cos(2\phi_f+4\tau)$ & ${1\over b_0^{5/2}}\left({3e_{p,0}^2\over 16}\right)$ \\
\hline
\end{tabular}\\
\end{center}

\vspace{2cm}

\noindent
$<{\cal F}^{(1)}>$\\
\begin{center}
\begin{tabular}{||l c || l c ||}
\hline
$\cos(\phi)$ & $+{1\over b_0^{3/2}}\left({3e_{p,0} e'\over 2}\right)+{1\over b_0^{5/2}}\left({-15e_{p,0} e'\over 8}\right)$ & $\cos(\tau-\phi)$ & ${e_{p,0} e'\over 4}+{1\over b_0^{3/2}}\left({-e_{p,0} e'\over 4}\right)+{1\over b_0^{5/2}}\left({-3e_{p,0} e'\over 2}\right)$ \\
\hline
$\cos(\tau+\phi)$ & ${e_{p,0} e'\over 4}+{1\over b_0^{3/2}}\left({-e_{p,0} e'\over 4}\right)+{1\over b_0^{5/2}}\left({21e_{p,0} e'\over 8}\right)$ & $\cos(2\tau-\phi)$ & $-e_{p,0} e'+{1\over b_0^{3/2}}\left({e_{p,0} e'\over 4}\right)+{1\over b_0^{5/2}}\left({15e_{p,0} e'\over 16}\right)$ \\
\hline
$\cos(2\tau+\phi)$ & $+{1\over b_0^{5/2}}\left(-{9e_{p,0} e'\over 16}\right)$ & $\cos(3\tau-\phi)$ & $+{1\over b_0^{5/2}}\left({3e_{p,0} e'\over 8}\right)$ \\
\hline
$\sin(\phi)$ & $+{1\over b_0^{3/2}}\left(-{3\sqrt{3}e_{p,0} e'\over 4}\right)+{1\over b_0^{5/2}}\left({21\sqrt{3}e_{p,0} e'\over 8}\right)$ & $\sin(\tau-\phi)$ & $-{\sqrt{3} e_{p,0} e'\over 4}+{1\over b_0^{3/2}}\left({\sqrt{3}e_{p,0} e'\over 4}\right)+{1\over b_0^{5/2}}\left({3\sqrt{3}e_{p,0} e'\over 4}\right)$ \\
\hline
$\sin(\tau+\phi)$ & ${\sqrt{3} e_{p,0} e'\over 4}+{1\over b_0^{3/2}}\left(-{\sqrt{3}e_{p,0} e'\over 4}\right)+{1\over b_0^{5/2}}\left(-{3\sqrt{3}e_{p,0} e'\over 4}\right)$ & $\sin(2\tau-\phi)$ & $+{1\over b_0^{5/2}}\left({9\sqrt{3}e_{p,0} e'\over 16}\right)$\\
\hline
$\sin(2\tau+\phi)$ & $+{1\over b_0^{5/2}}\left(-{9\sqrt{3}e_{p,0} e'\over 16}\right)$& $\qquad$ & $\qquad$ \\ 
\hline
\end{tabular}\\
\end{center}

\vspace{2cm}

\endgroup
\end{adjustwidth}

\newpage

\noindent
$\tilde{{\cal F}}^{(1)} = {\cal F}^{(1)}-<{\cal F}^{(1)}>$\\
\begin{adjustwidth}{-5em}{+3em}
\begingroup
\fontsize{9pt}{12pt}\selectfont
\begin{center}
\begin{tabular}{||l c || l c ||}
\hline
$\cos(\phi_f-\tau+\phi)$ & $-2e'+{1\over b_0^{3/2}}\left({e'\over 2}\right)$ & $\cos(\phi_f+\phi)$ & ${3e'\over 4}+{1\over b_0^{3/2}}\left({e'\over 4}\right)$ \\
\hline
$\cos(\phi_f+\tau+\phi)$ & $-{e'\over 2}+{1\over b_0^{3/2}}\left(-e'\right)$ &$\cos(\phi_f+2\tau+\phi)$ & $-{e'\over 4}+{1\over b_0^{3/2}}\left({e'\over 4}\right)$ \\
\hline
$\cos(2\phi_f-2\tau+2\phi)$ & $+{1\over b_0^{5/2}}\left({3e'^2\over 16}\right)$ & $\cos(2\phi_f-\tau+\phi)$ & $+{1\over b_0^{5/2}}\left(-{9e_{p,0} e'\over 8}\right)$ \\
\hline
$\cos(2\phi_f-\tau+2\phi)$ & $-{27e'^2\over 8}+{1\over b_0^{3/2}}\left({3e'^2\over 8}\right)+{1\over b_0^{5/2}}\left({3e'^2\over 16}\right)$ & $\cos(2\phi_f+\phi)$ & $3 e_{p,0} e'+{1\over b_0^{3/2}}\left(-{3e_{p,0} e'\over 4}\right)+{1\over b_0^{5/2}}\left({3e_{p,0} e'\over 16}\right)$ \\
\hline
$\cos(2\phi_f+2\phi)$ & ${3e'^2\over 2}+{1\over b_0^{3/2}}\left(-{e'^2\over 8}\right)+{1\over b_0^{5/2}}\left(-{63e'^2\over 32}\right)$ & $\cos(2\phi_f+\tau+\phi)$ & $-{e_{p,0} e'\over 8}+{1\over b_0^{3/2}}\left({e_{p,0} e'\over 8}\right)+{1\over b_0^{5/2}}\left(3 e_{p,0} e'\right)$ \\
\hline
$\cos(2\phi_f+\tau+2\phi)$ & $-{e'^2\over 16}+{1\over b_0^{3/2}}\left({e'^2\over 16}\right)+{1\over b_0^{5/2}}\left({3e'^2\over 2}\right)$ & $\cos(2\phi_f+2\tau+\phi)$ & $-e_{p,0} e'+{1\over b_0^{3/2}}\left(-{e_{p,0} e'\over 2}\right)+{1\over b_0^{5/2}}\left({-15e_{p,0} e'\over 8}\right)$ \\
\hline
$\cos(2\phi_f+2\tau+2\phi)$ & ${e'^2\over 2}+{1\over b_0^{3/2}}\left(-{e'^2\over 2}\right)+{1\over b_0^{5/2}}\left({9e'^2\over 8}\right)$ & $\cos(2\phi_f+3\tau+\phi)$ & $-{3e_{p,0} e'\over 8}+{1\over b_0^{3/2}}\left({3e_{p,0} e'\over 8}\right)+{1\over b_0^{5/2}}\left(-{3 e_{p,0} e'\over 8}\right)$ \\
\hline
$\cos(2\phi_f+3\tau+2\phi)$ & ${3e'^2\over 16}+{1\over b_0^{3/2}}\left(-{3e'^2\over 16}\right)+{1\over b_0^{5/2}}\left(-{15e'^2\over 16}\right)$ & $\cos(2\phi_f+4\tau+\phi)$ & $+{1\over b_0^{5/2}}\left({3 e_{p,0} e'\over 16}\right)$ \\
\hline
$\cos(2\phi_f+4\tau+2\phi)$ & $+{1\over b_0^{5/2}}\left(-{3e'^2\over 32}\right)$ &$\sin(\phi_f+\phi)$ & ${3\sqrt{3}e'\over 4}+{1\over b_0^{3/2}}\left(-{3\sqrt{3}e'\over 4}\right)$ \\
\hline
$\sin(\phi_f+\tau+\phi)$ & $-{\sqrt{3}e'\over 2}+{1\over b_0^{3/2}}\left({\sqrt{3}e'\over 2}\right)$ & $\sin(\phi_f+2\tau+\phi)$ & $-{\sqrt{3}e'\over 4}+{1\over b_0^{3/2}}\left({\sqrt{3}e'\over 4}\right)$ \\
\hline
$\sin(2\phi_f-\tau+2\phi)$ & $+{1\over b_0^{5/2}}\left(-{9\sqrt{3}e'^2\over 16}\right)$ & $\sin(2\phi_f+\phi)$ & $+{1\over b_0^{5/2}}\left({27\sqrt{3}e_{p,0} e'\over 16}\right)$ \\
\hline
$\sin(2\phi_f+2\phi)$ & ${3\sqrt{3}e'^2\over 2}+{1\over b_0^{3/2}}\left(-{3\sqrt{3}e'^2\over 8}\right)+{1\over b_0^{5/2}}\left({3\sqrt{3}e'^2\over 32}\right)$ &$\sin(2\phi_f+\tau+\phi)$ & $-{\sqrt{3}e_{p,0} e'\over 8}+{1\over b_0^{3/2}}\left({\sqrt{3}e_{p,0} e'\over 8}\right)+{1\over b_0^{5/2}}\left(-{9\sqrt{3}e_{p,0} e'\over 4}\right)$ \\
\hline
$\sin(2\phi_f+\tau+2\phi)$ & $-{\sqrt{3}e'^2\over 16}+{1\over b_0^{3/2}}\left({\sqrt{3}e'^2\over 16}\right)+{1\over b_0^{5/2}}\left({3\sqrt{3}e'^2\over 2}\right)$ &$\sin(2\phi_f+2\tau+\phi)$ & $-\sqrt{3}e_{p,0} e'+{1\over b_0^{3/2}}\left({\sqrt{3}e_{p,0} e'\over 4}\right)+{1\over b_0^{5/2}}\left(-{3\sqrt{3}e_{p,0} e'\over 8}\right)$\\
\hline
$\sin(2\phi_f+2\tau+2\phi)$ & $-{\sqrt{3}e'^2\over 2}+{1\over b_0^{3/2}}\left(-{\sqrt{3}e'^2\over 4}\right)+{1\over b_0^{5/2}}\left(-{15\sqrt{3}e'^2\over 16}\right)$ & $\sin(2\phi_f+3\tau+\phi)$ & $-{3\sqrt{3}e_{p,0} e'\over 8}+{1\over b_0^{3/2}}\left({3\sqrt{3}e_{p,0} e'\over 8}\right)+{1\over b_0^{5/2}}\left({3\sqrt{3}e_{p,0} e'\over 4}\right)$ \\
\hline
$\sin(2\phi_f+3\tau+2\phi)$ & $-{3\sqrt{3}e'^2\over 16}+{1\over b_0^{3/2}}\left({3\sqrt{3}e'^2\over 16}\right)+{1\over b_0^{5/2}}\left(-{3\sqrt{3}e'^2\over 16}\right)$ & $\sin(2\phi_f+4\tau+\phi)$ & ${1\over b_0^{5/2}}\left({3\sqrt{3}e_{p,0} e'\over 16}\right)$ \\
\hline
$\sin(2\phi_f+4\tau+2\phi)$ & $ {1\over b_0^{5/2}}\left({3\sqrt{3}e'^2\over 32}\right)$ & $\qquad$ & $\qquad$ \\
\hline
\end{tabular}
\end{center}
\endgroup
\end{adjustwidth}


\begin{thebibliography}{}

\bibitem{}
Arnold, V. I.: Instability of dynamical systems with several degrees 
of freedom,
Sov. Math. Dokt. {\bf 5}, 581-585 
(1964)

\bibitem{}
Beaug\'{e}, C., and Roig, F.: A semianalytical model for the motion of 
the Trojan asteroids: proper elements and families, Icarus {\bf 153}, 
391-415  (2001)

\bibitem{}
Beaug\'{e}, C., S\'{a}ndor, Z., \'{E}rdi, B., and S\"{u}li, A.: Co-orbital 
terrestrial planets in exoplanetary systems: a formation scenario,  
Astron. Astrophys. {\bf 463}, 359-367 (2007)

\bibitem{}
Bien, R., and Schubart, J.: Three characteristic orbital parameters for 
the Trojan group of asteroids,  Astron. Astrophys. {\bf 175}, 292-298
(1987)

\bibitem{}
Brown, E. W., Shook, C. A.: Planetary Theory. New York, 
Cambridge University Press, p. 256 
(1964)

\bibitem{}
Celletti, A., Giorgilli, A.: On the stability of the Lagrangian points in 
the spatial restricted problem of three bodies, Celest. Mech. Dyn. Astron. 
{\bf 50}, 31-58 
(1991)

\bibitem{}
Chirikov, B. V., Lieberman, M. A., Shepelyansky, D. L., and Vivaldi, F. M.: 
A theory of modulational diffusion, Physica D {\bf 14}, 289-304
(1985) 

\bibitem{}
Cresswell, P., and Nelson, R. P.: On the growth and stability of Trojan 
planets, Astron. Astrophys. {\bf 493}, 1141-1147 
(2009)

\bibitem{}
Deprit, A.: Limiting orbits around the equilateral centers of libration, 
Astron. J. {\bf 71}, 77-87
(1966)

\bibitem{}
Deprit, A., and Rabe, E.: Periodic trojan orbits for the resonances 
$1/12$, Astron. J. {\bf 74}, 317-320
(1968)

\bibitem{}
Deprit, A., and Price, J.F.: L'espace de phase autour de $L_4$ pour 
la r\'esonance interne $1/3$, Astron. Astrophys. {\bf 1}, 427-430
(1969)

\bibitem{}
Deprit, A., and Henrard, J.: Sur les orbites p\'{e}riodiques issues 
de $L_4$ \`{a} la r\'{e}sonance interne $1/4$, Astron. Astrophys. 
{\bf 3}, 88-93
(1969)

\bibitem{}
Ding, M., Bountis, T., and Ott, E.: Algebraic escape in higher 
dimensional Hamiltonian systems, Phys. Lett. A {\bf 151}, 395-400
(1990)

\bibitem{}
Dobrovolskis, A.: Effects of Trojan exoplanets on the reflex 
motions of their parent stars, Icarus {\bf 226}, 1636-1641
(2013)

\bibitem{}
Dullin, H. R., and Worthington, J. I.: The vanishing twist in 
the restricted three body problem, arXiv:1309.1280 [math-ph]
(2013)

\bibitem{}
Dvorak, R., Pilat-Lohinger, E., Schwarz, R., and Freistetter, 
F.: Extrasolar Trojan planets close to habitable zones, Astron. 
Astrophys. {\bf 426}, 37-40 
(2004)

\bibitem{}
Efthymiopoulos C., Contopoulos G., Voglis, N., and Dvorak, R.: 
Stickiness and Cantori, J. Phys. A Math. Gen. 30, 8167-8186  
(1997)

\bibitem{}
Efthymiopoulos, C., S\'{a}ndor Z.: Optimized Nekhoroshev estimates 
for the Trojan asteroids with a symplectic mapping model of co-orbital 
motion, MNRAS {\bf 364}, 253-271 
(2005)

\bibitem{}
Efthymiopoulos, C.: High order normal form stability estimates for 
co-orbital motion, Celest. Mech. Dyn. Astron. {\bf 117}, 101-112
(2013)

\bibitem{}
Efthymiopoulos, C., and P\'{a}ez, R. I.: \emph{in preparation} 
(2014)

\bibitem{}
\'{E}rdi, B.: Long periodic perturbations of Trojan asteriods, 
Celest. Mech. Dyn. Astron. {\bf 43}, 303-308
(1988)

\bibitem{}
\'{E}rdi, B.: The Trojan problem, Celest. Mech. Dyn. Astron. 
{\bf 65}, 149-164
(1997)

\bibitem{}
Erdi B., S\'{a}ndor Zs.: Stability of co-orbital motion in exoplanetary 
systems, Celest. Mech. Dyn. Astron. {\bf 92}, 113-121 
(2005)

\bibitem{}
\'{E}rdi B., Nagy I, S\'{a}ndor Zs., S\"{u}li, A., Fr\"{o}hlich G.: 
Secondary resonances of co-orbital motions, MNRAS {\bf 381}, 33-40 
(2007)

\bibitem{}
\'{E}rdi, B., Forg\'{a}cs-Dajka, E., Nagy, I., Rajnai, R.: A parametric 
study of stability and resonances around $L_4$ in the elliptic restricted 
three-body problem, Celest. Mech. Dyn. Astron. {\bf 104}, 145-158
(2009)

\bibitem{}
Froeschlé, C., Guzzo, M., Lega, E.: Graphical evolution of the 
Arnold web: from order to chaos, Science {\bf 289}, 2108-2110 
(2000)

\bibitem{}
Funk, B., Schwarz., R., S\"{u}li, A., and \'{E}rdi, B.: On the stability 
of possible Trojan planets in the habitable zone: an application to the 
systems HB 147513 and HD 210277, MNRAS {\bf 423}, 3074-2082
(2012)

\bibitem{}
Gabern, F., Jorba, A., and Locatelli, U.: On the construction of the 
Kolmogorov normal form for the Trojan asteroids, Nonlinearity {\bf 18}, 
1705-1734 
(2005)

\bibitem{}
Garfinkel, B.: Theory of the Trojan asteriods, Part I, Astron. J. 
{\bf 85}, 368-379
(1977)

\bibitem{}
Giorgilli, A., and Skokos, Ch.: On the stability of the Trojan asteroids, 
Astron. Astrophys {\bf 317}, 254-261 
(1997)

\bibitem{}
Giuppone, C. A., Ben\'{i}tez-Llambay, P., and Beaug\'{e}, C.: Origin and 
detectability of coorbital planets from radial velocity data, MNRAS 
{\bf 421}, 356-368
(2012)

\bibitem{}
Guzzo, M., and Lega, E.: Evolution of the tangent vectors and localization
of the stable and unstable manifolds of hyperbolic orbits by Fast Lyapunov 
Indicators, ArXiv 1307.6731 
(2013)

\bibitem{}
Haghighipour, N., Capen, S., Hinse, T.: Detection of Earth-mass and 
super-Earth Trojan planets using transit timing variation method, 
Cel. Mech. Dyn. Astron. {\bf 117}, 75-89
(2013)

\bibitem{}
Kov\'{a}cs, T., and Erdi, B.: Transient chaos in the Sitnikov problem, 
Cel. Mech. Dyn. Astron. {\bf 105}, 289-304
(2009)

\bibitem{}
Lai, Y. C., Ding, M., Grebogi, C., Bl\"{u}mel, R.: Algebraic decay and 
fluctuations of the decay exponent in Hamiltonian systems, Phys. Rec. A 
{\bf 46}, 4661-4669
(1992)

\bibitem{}
Laskar, J.: The chaotic motion of the solar system: A numerical estimate
for the size od the chaotic zones, Icarus {\bf 88}, 266-291
(1990)

\bibitem{}
Laskar, J., Joutel, F., and Boudin, F.: Orbital, precessional, and 
insolation quantities for the Earth from $-$20Myr to $+$10Myr, Astron. 
Astrophys. {\bf 270}, 522-533
(1993)

\bibitem{}
Laughlin, G., and Chambers, J.: Extrasolar Trojans: the viability and 
detectability of planets in the 1:1 resonance, Astroph. J. {\bf 124}, 
592-600
(2002)

\bibitem{}
Levison, H., Shoemaker, E. M., Shoemaker, C. S.: Dynamical evolution of 
Jupiter's Trojan asteroids, Nature {\bf 385}, 42-44 
(1997)

\bibitem{}
Libert, A.S., and Sansottera, M.: On the extension of the Laplace-Lagrange 
secular theory to order two in the masses for extrasolar systems, 
Cel. Mech. Dyn. Astron. {\bf 117}, 149-168
(2013)

\bibitem{}
Lohinger, E., and Dvorak, R.: Stability regions around $L_4$ in the elliptic
restricted problem, Astron. Astrophys. {\bf 280}, 683-687
(1993)

\bibitem{}
Lhotka, Ch., Efthymiopoulos C., and Dvorak, R.: Nekhoroshev stability at 
$L_4$ or $L_5$ in the elliptic-restricted three-body problem - application 
to Trojan asteroids, MNRAS {\bf 384}, 1165-1177
(2008)

\bibitem{}
Lyra, W., Johansen, A., Klahr, H., and Piskunov, N.: Standing on the 
shoulders of giants. Trojan Earths and vortex trapping in low mass 
self-gravitating protoplanetary disks of gas and solids,
Astron. Astrophys. {\bf 493}, 1125-1139
(2009)

\bibitem{}
Marzari, F., Tricarino P., and Scholl, H.: Stability of Jupiter Trojans 
investigated using frequency map analysis: the MATROS project, MNRAS 
{\bf 345}, 1091-1100
(2003)

\bibitem{}
Meiss, J. D.: Symplectic maps, variation principles and transport, 
Rev. Mod. Phys. {\bf 64}, 795-848
(1992)

\bibitem{}
Menou, K., and Tabachnik, S.: Dynamical habitability of known extrasolar 
planetary systems, Astroph. J. {\bf 583}, 473-488 
(2003)

\bibitem{}
Milani A.: The Trojan asteroid belt: Proper elements, stability, 
chaos and families, Celest. Mech. Dyn. Astron. {\bf 57}, 59-94 
(1993)

\bibitem{}
Morbidelli, A.: Modern Celestial Mechanics. Aspects of Solar System 
Dynamics, Taylor and Francis, London
(2002)

\bibitem{}
Morais, M. H. M.: Hamiltonian formulation of the secular theory for 
Trojan-type motion, Astron. Astrophys. {\bf 369}, 677-689
(2001)

\bibitem{}
Murray, C.D., Dermott, S.F.: Solar System Dynamics, 
Cambridge University Press.
(1999)

\bibitem{}
Namouni, F., and Murray, C.D.: The effect of eccentricity and inclination 
on the motion near the lagrangian points $L_4$ and $L_5$, Celest. Mech. 
Dyn. Astron. {\bf 76}, 131-138
(2000)

\bibitem{}
Rabe, E.: Third-order stability of the Long-period Trojan librations, 
Astron. J. {\bf 72}, 10-17
(1967)

\bibitem{}
Robutel, P., Gabern, F., Jorba, A.: The observed Trojans and the global 
dynamics around the lagrangian points of the Sun-Jupiter system, Celest. 
Mech. Dyn. Astron. {\bf 92}, 53-69 
(2005)

\bibitem{}
Robutel, P., Gabern, F.: The resonant structure of Jupiter's Trojan 
asteroids - I. Long-term stability and diffusion, MNRAS {\bf 372}, 
1463-1482 
(2006)

\bibitem{}
Robutel, P., Bodossian, J.: The resonant structure of Jupiter's Trojan 
asteroids - II. What happens for different configurations of the planetary 
systems, MNRAS {\bf 399}, 69-87 
(2009)

\bibitem{}
Schwarz, R., Dvorak, R., S\"{u}li, A, and \'{E}rdi B.: Survey of the 
stability region of hypothetical habitable Trojan planets, Astron. 
Astrophys. {\bf 474}, 1023-11029 
(2007)

\bibitem{}
Tsiganis, K., Varvoglis, H., Dvorak, R.: Chaotic diffusion and effective 
detectability of Jupiter Trojans, Celest. Mech. Dyn. Astron. {\bf 92}, 
71-87 
(2005)

\end{thebibliography}
\end{document}